\documentclass[epsfig,11pt]{article}
\usepackage{amsfonts}
\usepackage{amssymb}
\usepackage{graphicx}
\usepackage{amsmath}

\input epsf.sty
\newtheorem{theorem}{Theorem}
\newtheorem{acknowledgement}[theorem]{Acknowledgement}

\input{tcilatex}
\makeatletter \@addtoreset{equation}{section}
\renewcommand{\theequation}{\thesection.\arabic{equation}}
\begin{document}

\title{\rightline{\mbox{\small
{Lab/UFR-HEP0614/GNPHE/0614/VACBT/0614}}} \textbf{Black Holes in type IIA
String on  Calabi-Yau Threefolds} \textbf{with Affine ADE Geometries and
q-Deformed 2d Quiver Gauge Theories}}
\author{R. Ahl Laamara$^{{\small 1,2}}$, A. Belhaj$^{{\small 1,2,3,4}}$, L.B.
Drissi$^{{\small 1,2}}$, E.H. Saidi$^{{\small 1,2,3,5}}$ \\
{\small 1. Lab/UFR- Physique des Hautes Energies\thanks{%
ahllaamara@gmail, lbdrissi@gmail.com, h-saidi@fsr.ac.ma}, Facult\'{e} des
Sciences, Rabat, Morocco,}\\
{\small 2. GNPHE, Groupement National de Physique des Hautes Energies, Si%
\`{e}ge focal: FS, Rabat.}\\
{\small 3. Virtual African Centre for Basic Science \& Technology, Focal
point, Lab/UFR-PHE, FSR.}\\
{\small 4. Departamento de Fisica Teorica, Universidad de Zaragoza\thanks{%
belhaj@unizar.es}, 50009- Zaragoza, Spain.}\\
{\small 5. Academie Hassan II des Sciences \& Techniques, Coll\`{e}ge
Physique-Chimie, Royaume du Maroc.}}
\maketitle

\begin{abstract}
Motivated by studies on 4d black holes and q-deformed 2d Yang Mills theory,
and borrowing ideas from compact geometry of the blowing up of affine ADE
singularities, we build a class of local Calabi-Yau threefolds (CY$^{3}$)
extending the local 2-torus model $\mathcal{O}(m)\oplus \mathcal{O}(-m)\rightarrow T^{2\text{ }}$
considered in hep-th/0406058 to test OSV conjecture. We first study toric
realizations of $T^{2}$ and then build a toric representation of $X_{3}$
using intersections of local Calabi-Yau threefolds $\mathcal{O}(m)\oplus
\mathcal{O}(-m-2)\rightarrow \mathbb{P}^{1}$. 
We develop the 2d $\mathcal{N}=2$ linear $\sigma $-model for this
 class of toric CY$^{3}$s. Then we use these local
backgrounds to study partition function of 4d black holes in type IIA string theory 
and the underlying q-deformed 2d quiver gauge theories. We also make
comments on 4d black holes obtained from D-branes wrapping cycles in $%
\mathcal{O}\left( \mathbf{m}\right) \oplus \mathcal{O}\left( \mathbf{-m-2}%
\right) \rightarrow \mathcal{B}_{k}$ with $\mathbf{m=}\left(
m_{1},...,m_{k}\right) $ a $k$-dim integer vector and $\mathcal{B}_{k}$ a
compact complex one dimension base consisting of the intersection of $k$
2-spheres $S_{i}^{2}$ with generic intersection matrix $I_{ij}$. We give as
well the explicit expression of the q-deformed path integral measure of the
partition function of the 2d quiver gauge theory in terms of $I_{ij}.$
\newline
\textbf{Key words}: {\small Black holes in string theory, OSV conjecture,
q-deformed 2d quiver gauge theory, topological string theory}.
\end{abstract}
\newpage
\tableofcontents

\newpage

\section{Introduction}

\qquad 
Few years ago, Ooguri, Strominger and Vafa (OSV) have made a
conjecture \textrm{\cite{01}} relating the microstates counting of 4d BPS
black holes in type II string theory on Calabi-Yau threefolds $X_{3}$ to the
topological string partition function $Z_{top}$ on the same manifold. The
equivalence between the partition function $Z_{brane}$ of large $N$  D-branes,
and that of the associated 4d BPS black hole $Z_{BH}$ leads to the
correspondence $Z_{brane}=|Z_{top}|^{2}$ to all orders in $\frac{1}{N}$
expansion. OSV conjecture has brought important developments on this link:
it provides the non-perturbative completion of the topological string theor 
\textrm{\cite{02}-\cite{77} }and\textrm{\ }gives a way to compute the
corrections to 4d $\mathcal{N}=2$ Bekenstein-Hawking entropy \textrm{\cite%
{07,08}}. OSV relation has been extended in \textrm{\cite{05,P}} to open
topological strings, which capture BPS states data information on D-branes
wrapped on Lagrangian submanifolds of the Calabi-Yau 3-folds.

\qquad Evidence for OSV proposal has been obtained by using local Calabi-Yau
threefolds and some known results on 2d $U\left( N\right) $ Yang-Mills
theory \textrm{\cite{gt}}. It has been first tested in \textrm{\cite{02} }by
considering configurations of D-branes wrapped cycles\textrm{\ }$\mathcal{O}%
(m)\oplus \mathcal{O}(-m)\rightarrow T^{2}$ for $m$ a positive definite
integer.\textrm{\ }Then it has been checked in\textrm{\ \cite{2} }by\textrm{%
\ }using wrapped D-branes in a vector bundle of\textrm{\ }rank 2 non trivial
fibration over a genus g-Riemann surface $\mathcal{O}(2g+m-2)\oplus \mathcal{%
O}(-m)\rightarrow \Sigma _{g}$. It has been shown in these studies that the
BPS black hole partition function localizes onto field configuration which
are invariant under $U\left( 1\right) $ actions of the fibers. In this way,
the 4d gauge theory reduces effectively to q-deformed Yang-Mills theory on $%
\Sigma _{g}$. These works have been recently enlarged in \textrm{\cite{8}}
by considering local Calabi-Yau manifolds with torus symmetries such as
local $\mathbb{P}^{2}$=$\mathcal{O}(-3)\rightarrow \mathbb{P}^{2}$, local $F_{0}=%
\mathcal{O}(-2,-2)\rightarrow \mathbb{P}_{B}^{1}\times \mathbb{P}_{F}^{1}$
and $ALE\times C$. In the last example, the $A_{k}$ type ALE space is given
by gluing together $\left( k+1\right) $ copies of $C^{2}$ viewed as a real
two dimensional base fibered by torus $T^{2}$. Other related works have been
developed in \textrm{\cite{szabo1,szabo2,szabo3,szabo4,szabo5,m,n,9,10}}. M-theory and $AdS_{3}/CFT_{2}$
interpretations of OSV formula have been also studied in \textrm{\cite{3}.}

\qquad In this paper we contribute to the program of testing OSV conjecture
for massive 4d black holes on toric Calabi-Yau threefolds in connection with
q-deformed quiver gauge theories in two dimensions. This study has been
motivated by looking for a 2d $\mathcal{N}=2$ supersymmetric gauged linear
sigma model of $\mathcal{O}(m)\oplus \mathcal{O}(-m)\rightarrow T^{2}$. More
precisely we consider a special class of local Calabi-Yau threefolds which
combine features of both $\mathcal{O}(m)\oplus \mathcal{O}(-m)\rightarrow
T^{2}$ and $\mathcal{O}(m)\oplus \mathcal{O}(-m-2)\rightarrow S^{2}$, and
involves toric graphs that look like affine ADE Dynkin diagrams and beyond.
Recall that affine ADE\ geometries are known to be described by elliptic
fibration over $C^{2}$ and lead to $\mathcal{N}=2$ conformal quiver gauge
theories in 4d space-time with gauge groups $G=\Pi _{i}U\left( s_{i}M\right)
$, where the positive integers $s_{i}$ are the Dynkin weights of affine
Kac-Moody algebras. Borrowing the method of the above quiver gauge theories
\textrm{\cite{5,6,7}} and using the results of \cite{02}, we engineer a new
class of local Calabi-Yau threefolds that enlarges further the class of CY3s
used before and that agrees with OSV conjecture.

\qquad Our study gives moreover an explicit toric representation of $%
\mathcal{O}(m)\oplus \mathcal{O}(-m)\rightarrow T^{2}$ especially if one
recalls that $T^{2}$ viewed as $S^{1}\times S^{1}$ does not have, to our
knowledge, a simple nor unique toric realization. As we know real skeleton
base of toric diagrams representing toric manifolds requires at least one
2-sphere $S^{2}$ which is not the case for the simplest $S^{1}\times S^{1}$
geometry. In the analysis to be developed in this study, the 2-torus will be
realized by using special linear combinations $\left[ \Delta _{n+1}\right] $
of intersecting 2-spheres with same homology as a 2-torus. The positive
integer $n\geq 1$ refers to the arbitrariness in the number of $\left[ S^{2}%
\right] $'s one can use to get the elliptic curve class $\left[ T^{2}\right]
$. Among our main results, we mention that 2d quiver gauge theories,
associated with BPS black holes in type IIA string on the local CY$^{3}$s we
have considered, are classified by the "sign" of the intersection matrix $%
I_{ik}$ of the real 2-spheres $S_{i}^{2}$ forming the compact base
\begin{equation*}
I_{ik}=\left[ S_{i}^{2}\right] \left[ S_{k}^{2}\right] ,\qquad i,k=0,...,n.
\end{equation*}
According to whether $\sum_{k}I_{ik}u_{k}>0$, $\sum_{k}I_{ik}u_{k}=0$ or $%
\sum_{k}I_{ik}u_{k}<0$ for some positive integer vector $\left( u_{k}\right)
$, we then distinguish three kinds of local models. For the second case
\begin{equation*}
\sum_{k}I_{ik}u_{k}=0,
\end{equation*}
the $u_{k}$'s are just the Dynkin weights  of affine Kac-Moody
algebras and the corresponding 2d quiver gauge theory is \textit{non-deformed%
} in agreement with the result of \cite{02}. The obove relation corresponds
then to the case where the genus g-Riemann surface is a 2-torus; i.e
\begin{equation*}
2g-2=0,\qquad \leftrightarrow \qquad g=1.
\end{equation*}%
For the two other cases, the gauge theory is \textit{q-deformed} and
recovers, as a particular case, the study of \cite{8} dealing with ALE
spaces.

\qquad On the other hand, this study will be done in the type IIA string theory  set
up. So we shall also use our construction to complete partial results in
literature on field theoretic realization using $2d$ $\mathcal{N}=\left(
2,2\right) $ linear sigma model for affine ADE geometries with special focus
on the $\widehat{A}_{n}$ case. To our knowledge, this supersymmetric 2d
field realization of local Calabi-Yau threefolds has not been considered
before.

\qquad The organization of paper is as follows: In section 2, we begin by
recalling general features on toric graphs. Then we study the toric
realizations of local $T^{2\text{ }}$ using the techniques of blowing up of
affine ADE geometries. We study also the field theoretic realization of the
non trivial fibrations of local $T^{2\text{ }}$ which corresponds to
implementing framing property \textrm{\cite{4}}. In section 3, we develop
the $\mathcal{N}=2$ supersymmetric gauged linear sigma model describing the
local torus geometry. This study gives an explicit field theoretic
realization of geometric objects such as the surface divisors of the local
2-torus and their edge boundaries in terms of field equations of motion and
vevs. In section 4, we construct the 4d BPS black holes in type IIA string
by considering brane configurations using D0-D2-D4-branes in the non
compact 4-cycles. We show, amongst others, that the gauge theory of the 
D-branes, which is dual to topological strings on the Calabi-Yau threefold,
localizes to a "q-deformed" 2d quiver gauge theory on the compact part of
the affine ADE geometry and test OSV conjecture. More precisely, we show
that the usual power $\left( 2g-2\right) $ of the weight of the deformed
path integral measure for $\mathcal{O}(2g+m-2)\oplus \mathcal{O}%
(-m)\rightarrow \Sigma _{g}$ gets replaced, in case of 2d quiver gauge
theories, by the intersection matrix $I_{ij}$. There, we show that the
property $\left( 2g-2\right) =0$ for $g=1$ corresponds to the identity $%
\sum_{j}I_{ij}s_{j}=0$ of affine Kac-Moody algebras. Motivated by this link,
we study 4d black holes based on D-branes wrapping cycles in $\mathcal{O}%
\left( \mathbf{m}\right) \oplus \mathcal{O}\left( \mathbf{-m-2}\right)
\rightarrow \mathcal{B}_{k}$ with $\mathbf{m=}\left( m_{1},...,m_{k}\right) $
an integer vector and where $\mathcal{B}_{k}$ is a complex one dimension
base consisting of the intersection of $k$ 2-spheres $S_{i}^{2}$ with generic
intersection matrix $I_{ij}$. In section 5, we give our conclusion and
outlook.

\section{Toric realization of local $T^{2\text{ }}$}

\qquad In this section, we build toric representations of the class of local
2-torus $\mathcal{O}(m)\oplus \mathcal{O}(-m)\rightarrow T^{2\text{ }}$ by developing the idea
outlined in the introduction. There, it was observed that although, strictly
speaking, $T^{2\text{ }}=S^{1}\times S^{1}$ is not a toric manifold (base
reduced to two points), it may nevertheless be realized by gluing several
2-spheres in very special ways. Before showing how this can be implemented
in the above local CY3, recall that the study of local threefold geometry
\begin{equation}
\mathcal{O}(m)\oplus \mathcal{O}(-m)\rightarrow T^{2\text{ }}  \label{1}
\end{equation}%
is important from several views. It has been used in \textrm{\cite{02}} to
test OSV conjecture \textrm{\cite{01}} and was behind the study of several
generalizations, in particular
\begin{equation}
\mathcal{O}(m+2g-2)\oplus \mathcal{O}(-m)\rightarrow \Sigma _{g}.  \label{2}
\end{equation}%
The novelty brought by these class of local CY3s stems also from the use of
the non trivial rank 2 fiber $\mathcal{O}(p)\oplus \mathcal{O}(-m)$, with $%
p=m+2g-2$ rather than $\mathcal{O}(2g-2)\oplus \mathcal{O}(0)$. These non
trivial fibers, which were motivated by implementing twisting by framing
\textrm{\cite{4}}, turn out to play a crucial role in the study of  4d BPS
black holes from type IIA string theory compactification. Generally, the
class of local CY3s eq(\ref{2}) which will be considered later (section
\textrm{4}), is mainly characterized by two integers $m$ and the genus $g$
and may be generically denoted as follows
\begin{equation}
X_{3}^{\left( k_{3},k_{2},k_{1}\right) },\qquad g=0,1,...;\qquad m\in Z.
\label{lc}
\end{equation}%
Here $k_{1}=2-2g,$ $k_{2}=-m$ and $k_{3}=m+2g-2$ satisfy the Calabi-Yau
condition $\sum_{i}k_{i}=0$ leaving only two free integers $m$ and $g$. By
making choices of these integers one picks up a particular local CY3. For $%
g=m=0$ and $g=1,$ $m=0$ for example, we have $\mathcal{O}(0)\oplus \mathcal{O%
}(-2)\rightarrow \mathbb{P}^{1}$ and $\mathcal{O}(0)\oplus \mathcal{O}%
(0)\rightarrow T^{2\text{ }}$ respectively. The local CY$^{3}$s (\ref{lc})
may be also viewed as a line bundle $\mathcal{L}_{\mathcal{D}_{\left(
m,g\right) }}^{\left( m+2g-2\right) }$ of the complex two dimensional\
divisor
\begin{equation}
\left[ \mathcal{D}_{\left( m,g\right) }\right] =\mathcal{O}(-m)\rightarrow
\Sigma _{g}.
\end{equation}%
This local complex surface $\left[ \mathcal{D}_{\left( m,g\right) }\right] $
has a compact curve $\Sigma _{g}$ with the following intersection number
\begin{equation}
\left[ \mathcal{D}_{\left( m,g\right) }\right] .\left[ \Sigma _{g}\right]
=m+2g-2.  \label{4d}
\end{equation}%
In the case where $\Sigma _{g}$ is a 2-torus ($g=1$), the above two integers
series of local CY$^{3}$ reduces to the one integer threefold series $%
X_{3}^{\left( m,-m,0\right) }$. The previous non compact real 4-cycles are
then given by \textrm{\cite{02}}%
\begin{equation}
\left[ \mathcal{D}_{\left( m,g\right) }\right] =\mathcal{O}(-m)\rightarrow
T^{2},  \label{4c}
\end{equation}%
and their intersection number with the 2-torus class $\left[ T^{2}\right] $
is $\left[ \mathcal{D}_{m,g}\right] .\left[ T^{2}\right] =m$.

\subsection{Toric realization}

\qquad Our main objectives in this subsection deals with the two following: (%
\textbf{1}) Build explicit toric realisations of the local 2-torus by using
particular realisations of the real 2-cycle $\left[ T^{2}\right] $. These
realisations, which will be used later on, have been motivated from results
on blowing up of affine ADE singularities of ALE spaces and geometric
engineering of 4d  $\mathcal{N}=2$ supper QFTs \textrm{\cite{5,6,7}}. It
gives a powerful tool for the explicit study of the special features of
local 2-torus and allows more insight in the building of new classes of
local Calabi-Yau threefolds for testing OSV conjecture.\newline
(\textbf{2}) Use the result of the analysis of point (1) to complete partial
results in literature on type IIA geometry with affine ADE singularities.
More precisely, we construct the $2d$ $\mathcal{N}=2$ supersymmetric gauged
linear sigma model
\begin{equation}
\mathcal{S}_{2d}^{\mathcal{N}=2}\sim \int d^{2}\sigma d^{4}\theta \left(
\sum_{i}\Phi _{i}^{+}e^{\sum_{a}q_{i}^{a}V_{a}}\Phi _{i}+\sum_{a}\xi
^{a}V_{a}\right) ,
\end{equation}%
giving the field realization of local 2-torii. This construction to be
developed further in section 3, will be used for the two following: \newline
(\textbf{a}) Work out explicitly the results of (q-deformed) 2d YM theories
on $T^{2}$ and give their extensions to 2d quiver gauge theories on the
elliptic curve realized a linear combination of intersecting 2-spheres.
\newline
(\textbf{b}) Study partition function properties of 4d BPS black holes along
the lines of \textrm{\cite{3}} and ulterior studies \textrm{\cite{8,m,n,9,10} }%
to test OSV conjecture.

\subsubsection{General on toric graphs}

\qquad In building toric realization of local CY$^{3}$s \cite{11,12,13,14,141}, one encounters few
basic objects that do almost the complete job in striking analogy with the
work done by Feynman graphs in perturbative QFTs. In particular, one has:
\newline
(\textbf{i}) "Propagators" given by the toric graph of the real 2-sphere $%
S^{2}$. It corresponds to the two points free field Green function in the
language of quantum feld theory (QFT).
\begin{figure}[tbph]
\begin{center}
\hspace{0cm} \includegraphics[width=8cm]{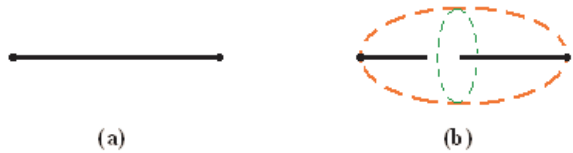}
\end{center}
\caption{\textbf{(a) }{\protect\small {Toric graph of a compact 2-sphere
with Kahler modulus }$\mathit{r}\neq 0$\textbf{{.} (b) }{The fattening of
compact 2-sphere where the circle S}$^{1}${\ is represented. S}$^{1}$ {%
shrinks at the fixed points of the U}$\left( 1\right) ${\ action. }}}
\end{figure}
Recall that the real 2-sphere $S^{2}$ is given by the compactification of
the complex line $C$. The latter can be realized (polar coordinates) as the
half line $R_{+}$ with fiber $S^{1}$ that shrinks at the origin. The
compactification of $C$; i.e $\mathbb{P}^{1}$ the complex one dimensional projective
space, is obtained by restricting $R_{+}$ to a finite straight line (a
segment) which is interpreted as a propagator in the language of QFT Feynman
graphs \textrm{\cite{4}}. Alike, we distinguish here also two situations:
\newline
($\mathbf{\alpha }$) Finite (internal) lines which are associated with
compact 2-spheres and are interpreted in terms of propagating closed string
states. \newline
($\mathbf{\beta }$) Infinite (external) lines with non compact 2-spheres
(discs/ complex plane) and are used to implement open string state
contributions to topological string amplitudes. These open string states end
on D-branes.\newline
With these propgators, one already build complex one and two dimensional  toric
manifolds as shown on figure below.
\begin{figure}[tbph]
\begin{center}
\hspace{-1cm} \includegraphics[width=8cm]{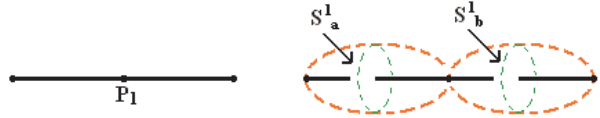}
\end{center}
\caption{{\protect\small {(a) }{Toric quiver of two intersecting 2-spheres}%
\textbf{, (b) }{\ the corresponding fat toric graph where we have also
represented the circles associated with the two U}$\left( 1\right) ${\ toric
actions. }}}
\end{figure}
To construct CY$^{3}$, we need 3-vertices which we discuss in the next.
\newline
(\textbf{ii}) The 3-vertex can be thought of locally as the intersection of
the three complex lines of $C^{3}$ \textrm{\cite{4}}. This vertex plays a
crucial role in building local CY$^{3}$s. For instance $\mathcal{O}\left(
-3\right) \rightarrow \mathbb{P}^{2}$ has three the 3-vertices corresponding
to the three fixed points of the $U^{4}\left( 1\right) /U\left( 1\right) $
toric actions. The edge propagators (2-spheres) are fixed under $U^{2}\left(
1\right) $ subgroups of $U^{4}\left( 1\right) /U\left( 1\right) $

\begin{figure}[tbph]
\begin{center}
\hspace{-1cm} \includegraphics[width=8cm]{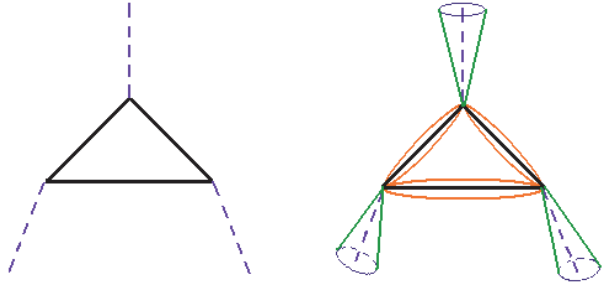}
\end{center}
\caption{{\protect\small {Toric graph of $ \mathcal{O}\left( -3\right) \rightarrow
\mathbb{P}^{2}$. The compact part is $\mathbb{P}^{2}$. It consists of three
intersecting $\mathbb{P}^{1}$'s and three vertices. (\textbf{a}) Figure (on
left) represents real skeleton. (\textbf{b}) Figure (on right) gives its
fattening.}}}
\end{figure}
With these objects one can build other local CY3s. Using fat propagators and
vertices we can also have a picture on their internal topology. One may also
describe its even dimensional homology cycles. Real 2-cycles $\mathcal{C}%
_{i} $ of local CY3s are represented by linear combinations of segments and
real 4-cycles $\mathcal{D}_{i}$ (divisors of CY$^{3}$s) by $2d$ polygons: a
triangle for $\mathbb{P}^{2}$, a rectangle for $\mathbb{P}_{fiber}^{1}\times%
\mathbb{P}_{base}^{1}$ and so on. Compact 4-cycles have then finite size.

The power of the toric quiver realization of threefolds comes also from its
simplicity due to the fact that the full structure of toric CY$^{3}$ quiver
diagrams is basically captured by the lines (2-cycles) since boundaries of
divisors (4-cycles) are given by taking cross products of pairs of straight
line generators.

Note that, though $\mathbb{P}^{1}$ is not a Calabi-Yau submanifold since its
first Chern class is $c_{1}\left( \mathbb{P}^{1}\right) =2$,  this one
dimensional complex projective space, together with the vertex, are the
basic objects in drawing the 2d graphs of toric manifolds. Note also that
torii $S^{1},$ $T^{2}$ and $T^{3}$ of CY$^{3}$ appear in this construction
as fibers and play a fundamental role in the study of topological string
theory amplitudes \textrm{\cite{4,15,16}}.

\subsubsection{Toric realisations of $X_{3}^{\left( m,-m,0\right) }$}

\qquad It has been shown in \textrm{\cite{02}} that the one integer series
of spaces $X_{3}^{\left( m,-m,0\right) }=$ $\mathcal{O}(m)\oplus \mathcal{O}%
(-m)\rightarrow $ $T^{2}$ is a toric local Calabi-Yau threefold used to
check OSV conjecture. From the above study on toric graphs it follows that,
a priori, one should be able to draw its corresponding toric quiver diagram.
However, unlike the 2-sphere, the usual 2-torus $S^{1}\times S^{1}$ has no
simple toric graph realization. Then the question is what kind of 2-torii $%
T^{2}$ are involved in $X_{3}^{\left( m,-m,0\right) }$? \newline
Here below, we would \ like to address this problem by using the graphic
method of toric geometry. In particular, we develop a way to build toric
graphs representating classes of $\left[ T^{2}\right] $. This will be done
by realizing $\left[ T^{2}\right] $ in terms of intersecting 2-spheres $%
S_{i}^{2}$ by expressing $\left[ T^{2}\right] $ as a "sum" of intersecting
2-spheres,%
\begin{equation*}
\left[ T^{2}\right] \sim \sum_{i}u_{i}\left[ S_{i}^{2}\right] ,\qquad
u_{i}\in \mathbb{Z}_{+},
\end{equation*}%
and thinking about $X_{3}^{\left( m,-m,0\right) }$ as a special limit of a
family of local CY3s
\begin{equation}
X_{3}^{\left( m,-m,0\right) }\quad \rightarrow \quad X_{3}^{\left( \mathbf{m}%
,-\mathbf{p},\mathbf{p-m}\right) }=\mathcal{O}(\mathbf{m})\oplus \mathcal{O}%
(-\mathbf{p})\rightarrow \Delta _{n+1}  \label{cg}
\end{equation}%
where $\mathbf{m}$ and $\mathbf{p}$ are some $\left( n+1\right) $-
dimensional integer vectors given by%
\begin{equation}
\mathbf{m}\mathbf{=}\left( m_{0},m_{0},...,m_{n}\right) ,\qquad \mathbf{p}%
\mathbf{=}\left( p_{0},p_{0},...,p_{n}\right).
\end{equation}%
In this way  the  Calabi-Yau condition reads
\begin{equation}
p_{i}-m_{i}=2,\qquad i=0,...,n,
\end{equation}%
which we denote formally as $\mathbf{p-m=2}$. Then, we extend the obtained
results to build the toric graphs describing $X_{3}^{\left( \mathbf{m},-%
\mathbf{p},\mathbf{2}\right) }$ and their even real homology cycles. Here it
is interesting to note the two following: \newline
(\textbf{a}) This construction is important since once we have the toric
quivers, one can use them for different purposes. For instance, we can use
the toric graphs in the topological vertex method of \textrm{\cite{4,16}} to
compute explicitly the partition functions of the topological string on $%
X_{3}^{\left( \mathbf{m},-\mathbf{p},\mathbf{2}\right) }$.\newline
(\textbf{b}) The local CY$^{3}$ we propose $X_{3}^{\left( \mathbf{m},-%
\mathbf{p},\mathbf{p-m}\right) }$ is not exactly $X_{3}^{\left(
m,-m,0\right) }$ considered in \textrm{\cite{02}; }it is more general. These
manifolds have different Kahler moduli spaces and their $U\left( 1\right)
\times U\left( 1\right) $ isometry groups are realized differently. To fix
the idea, think about the homogeneity group of the compact geometry $\Delta
_{n+1}$ of eq(\ref{cg}) as
\begin{equation}
U\left( 1\right) \times \frac{U^{n+1}\left( 1\right) }{U^{n}\left( 1\right) }%
\sim U\left( 1\right) \times U\left( 1\right) ,\qquad n>1.
\end{equation}%
To proceed we shall deal separately with base $\Delta _{n+1}$ and the two
line fibers of $X_{3}^{\left( \mathbf{m},-\mathbf{p},\mathbf{p-m}\right) }$.
We first look for toric quivers to describe the $\Delta _{n+1}$ class and $%
\mathcal{O}(\pm k_{0},..,\pm k_{n})$ independently. Then we use 3-vertex of $%
C^{3}$ and the Calabi-Yau condition to glue the various pieces. At the end,
we get the right real 3-dimensional toric graph of our local CY$^{3}$s and,
by fattening, the topology of $X_{3}^{\left( \mathbf{m},-\mathbf{p},\mathbf{2%
}\right) }$. This approach is interesting since one can control completely
the engineering of the toric quivers of local Calabi-Yau threefolds.
Moreover, seen that $\mathcal{O}(\pm k_{0},..,\pm k_{n})$ are line bundles,
their toric graphs are mainly the toric graphs of $\mathcal{O}(\pm m)$ which are
locally given by $C$ patches. What remains is the determination of toric
quiver of $\Delta _{n+1}$ class. In the figures given below,
\begin{figure}[tbph]
\begin{center}
\hspace{0cm} \includegraphics[width=6cm]{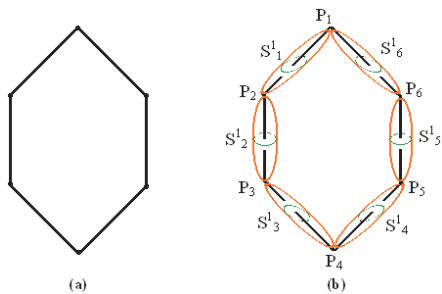}
\end{center}
\caption{{\protect\small {Figure (a) describes the toric graph for }$\Delta
_{6}${\ with six intersecting 2-spheres. Figure (b) gives its fattening
where the S}$_{i}^{1}${\ circles above the spheres S}$_{i}^{2}${\ and the
intersection points P}$_{i}${\ are also represented.}}}
\end{figure}
we develop two illustring examples. The first one (figure 4) concerns the
compact base $\Delta _{6}$ using six intersecting 2-spheres in the same
spirit one uses for the blowing up of singularities of ALE complex surfaces.
The second example ( figure 5) has in addition external non compact lines.
\begin{figure}[tbph]
\begin{center}
\hspace{0cm} \includegraphics[width=4cm]{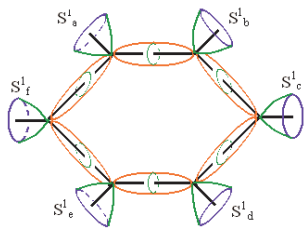}
\end{center}
\caption{{\protect\small {This figure represents the fattening of the toric
graph of }$\Delta _{6}${\ with 3-vertices and external legs ending on D
branes discs.}}}
\end{figure}
\newline
In what follows, we show that these toric graphs and their fattening
constitutes in fact particular topologies of infinitely many possible graphs.

\paragraph{(a) Solving the constraint equation of 2-torus homology: \newline
}

As noted before, the question of drawing a toric graph for $%
T^{2}=S^{1}\times S^{1}$ seems to have no sense at first sight. This is
because the basic (irreducible) real 2-cycle in toric geometry is $\mathbb{P}^{1}\sim
S^{2}$ with self intersection
\begin{equation}
\left[ \mathbb{P}^{1}\right] .\left[ \mathbb{P}^{1}\right] =-2.
\end{equation}%
Its toric graph is a straight line segment with length given by the size of
the 2-sphere (Kahler modulus $r$). The 2-torus has a zero self-intersection%
\begin{equation}
\left[ T^{2}\right] .\left[ T^{2}\right] =0,  \label{t2}
\end{equation}%
and a priori has no simple nor unique toric diagram. Tori $T^{n}$ are
generally speaking associated with $U^{n}\left( 1\right) $ phases of complex
variables. For instance, on complex line $C$ with local coordinate $z$, the
unit circle $S^{1}$ is given by $\left\vert z\right\vert =1$ and the $%
U\left( 1\right) $ symmetry acts as $z\rightarrow e^{i\theta }z$. This
circle and the associated $U\left( 1\right) $ symmetry are exhibited when
fattening toric graphs as shown on previous figures.\newline
To build a toric representation of $T^{2}$; but now viewed as 
$\Delta _{n+1}$, we use intersecting 2-spheres with particular combinations as
\begin{equation}
\Delta _{n+1}=\sum_{i=0}^{n}\epsilon _{i}\left[ \mathcal{C}^{i}\right]
,\qquad \mathcal{C}^{i}\sim \mathbb{P}_{i}^{1},  \label{t}
\end{equation}%
where the positive integers $\epsilon _{i}$ are obtained by solving eq(\ref%
{t2}). Denoting by $I_{ij}$ the intersection matrix of the 2-cycles $%
\mathcal{C}^{i}$,%
\begin{equation}
\left[ \mathcal{C}^{i}\right] .\left[ \mathcal{C}^{j}\right] =I^{ij},\qquad
I^{ii}=-2,
\end{equation}%
then the condition to fulfil eq(\ref{t2}) is
\begin{equation}
\sum_{i,j=0}^{n}\epsilon _{i}I^{ij}\epsilon _{j}=\sum_{i=0}^{n}\epsilon
_{i}\left( \sum_{j=0}^{n}I^{ij}\epsilon _{j}\right) =0,\qquad \epsilon
_{k}\in \mathbb{N}.
\end{equation}%
A solution of this constraint relation is given by taking
\begin{equation}
I^{ij}=-K_{ij}\left( \widehat{g}\right),
\end{equation}%
as minus the generalized Cartan matrix of affine Kac-Moody algebras $%
\widehat{g}$. In this case, the positive integers $\epsilon _{j}$ are
interpreted as the Dynkin weights and the topology of the 2-torus is same as
the affine Dynkin diagrams. In particular for the simplest case of simply
laced affine ADE are reported below.

\begin{figure}[tbph]
\begin{center}
\hspace{0cm} \includegraphics[width=7cm]{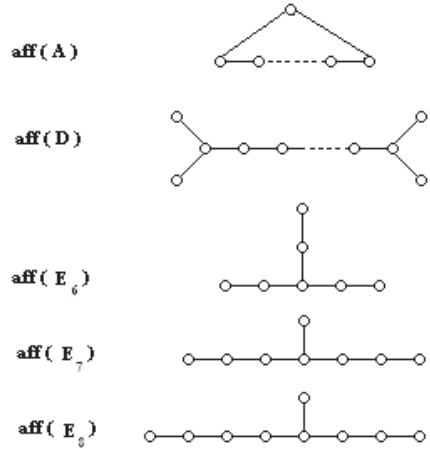}
\end{center}
\caption{{\protect\small {Affine simply laced ADE Dynkin diagrams. In
homology language, dots represent the 2-spheres and links the intersection%
\textbf{s}. In toric language, the Poincar\'{e} dual of these diagrams are
associated with toric graphs of 2-torii.}}}
\end{figure}
Therefore, there are infinitely many toric quivers realizing $T^{2}$ in
terms of intersecting 2-spheres. But to fix the ideas we will mainly focus
on the series based on affine $\widehat{A}_{n}$ Kac-Moody algebras. In this
case the elliptic curve is
\begin{equation}
\Delta _{n+1}=\sum_{i=0}^{n}\mathcal{C}^{i},\qquad \epsilon _{i}=1.
\label{el}
\end{equation}%
It involves $\left( n+1\right) $\ real 2-cycles with intersection matrix
\begin{eqnarray}
\mathcal{C}_{i}.\mathcal{C}_{j} &=&-\widehat{K}_{ij},\qquad i,\text{ \ }j=0,1,...,n,  \notag \\
\widehat{K}_{ij} &=&\delta _{i-1,j}-2\delta _{ij}+\delta _{i+j},\qquad
n+1\equiv 0.
\end{eqnarray}%
The curve $\Delta _{n+1}$ has the homology of a 2-torus realized by the
intersection of $\left( n+1\right) $ 2-spheres with the topology of Dynkin
diagram of affine $\hat{A}_{n}$.\newline
The Kahler parameter $r$ of the curve $\Delta _{n+1}$ defined as
\begin{equation}
r=\int_{\Delta _{n+1}}\mathbf{\omega ,}
\end{equation}%
with $\mathbf{\omega }$\ being the usual real Kahler 2-form, is given by the
sum over the kahler parameters $r^{i}$ of the 2-cycles $\mathcal{C}^{i}$
making up $\Delta _{n+1}$. We have
\begin{equation}
\mathbf{\omega }=\sum_{i}r^{i}\mathbf{\omega }_{i},\qquad \int_{\mathcal{C}%
^{i}}\mathbf{\omega }_{j}=\delta _{j}^{i}  \label{ef}
\end{equation}%
and so the Kahler modulus of $\Delta _{n+1}$ is given by 
\begin{equation}
r=\sum_{i=0}^{n}\int_{\mathcal{C}_{i}}\mathbf{\omega }_{i}=%
\sum_{i=0}^{n}r_{i}.
\end{equation}%
In this computation we have ignored the volume of the intersection points of
the 2-spheres $\mathcal{C}^{i}$ and $\mathcal{C}^{j}$ since they are
isolated points and moreover their volumes vanish in any case. To fix the
ideas, we shall set
\begin{equation}
r\geq r_{0}\geq r_{1}\geq ...\geq r_{n}\geq 0,  \label{r}
\end{equation}%
and for special computation, in particular when we study the path integral
of the partition function on quiver gauge theory on $\Delta _{n+1}$ dual to
4d black hole (section 4), we will in general sit at the moduli space point
where
\begin{equation}
r_{i}=\frac{r}{n+1}.
\end{equation}%
In all these cases, the Kahler moduli $r_{i}$ are positive. We shall also
suppose that we are away from $r=0$ describing the singularity of the curve $%
\Delta _{n+1}$ and where full non abelian gauge symmetry of quiver theory is
restored.

\paragraph{(\textbf{b}) 4-cycles of $\mathcal{O}(\mathbf{m})\oplus \mathcal{O}(-\mathbf{p}%
)\rightarrow \Delta _{n+1}$ \newline
}

Using the above realization, one can go ahead and build the 4-cycle and the
local Calabi-Yau threefold. Viewed as a whole, the non compact 4-cycle is
\begin{equation}
\mathcal{D}:\mathcal{O}(-\mathbf{p})\rightarrow \left( \sum_{i=0}^{n}\left[
S_{i}^{2}\right] \right) ,
\end{equation}%
with toric graph as given below.

\begin{figure}[tbph]
\begin{center}
\hspace{0cm} \includegraphics[width=8cm]{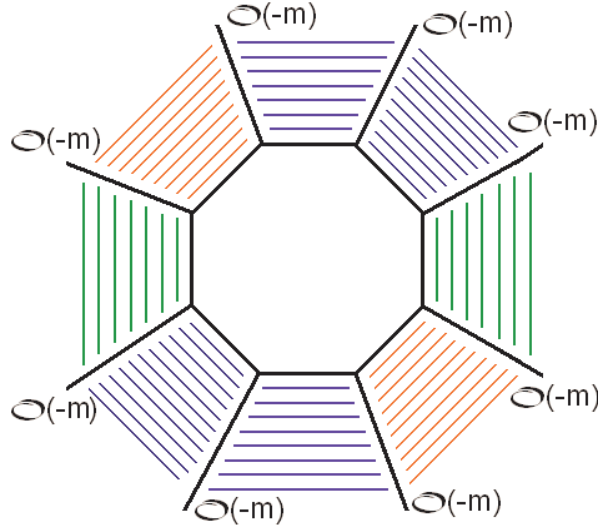}
\end{center}
\caption{Toric graph of a non compact 4-cycle of the local Calabi-Yau
Threefold. The compact part consists of eight intersecting 2-spheres.}
\end{figure}
Now using eq(\ref{t2}), the relation $\left[ \mathcal{D}\right] .\left[
\Delta _{n+1}\right] =m$ giving the intersection number which reads as
\begin{equation}
\left[ \mathcal{D}\right] .\left[ \sum_{i=0}^{n}\left[ S_{i}^{2}\right] %
\right] =m
\end{equation}%
and splitting $m$ as
\begin{equation}
m=\sum_{i=0}^{n}\left( m_{i}-2\right) ,\qquad p_{i}=m_{i}-2.  \label{sp}
\end{equation}%
It is not difficult to see that $\left[ \mathcal{D}\right] $ can be
decomposed as follows:
\begin{equation}
\left[ \mathcal{D}\right] =\sum_{i=0}^{n}\left[ \mathcal{D}_{i}\right]
,\qquad \mathcal{D}_{i}=\mathcal{O}(-p_{i})\rightarrow S_{i}^{2}  \label{ds}
\end{equation}%
with the property
\begin{equation}
\left[ \mathcal{D}\right] .\left[ S_{i}^{2}\right] =m_{i}-2.
\end{equation}%
If we take $m$ a positive integer, then we should have
\begin{equation}
\sum_{i=0}^{n}m_{i}>2\left( n+1\right)
\end{equation}%
as required by positivity of the intersection number.

\paragraph{(\textbf{c}) Toric graph of $X_{3}^{\left( \mathbf{m},-\mathbf{p},%
\mathbf{2}\right) }$ \ \newline
}

Similarly, we can build the toric graph of the local Calabi-Yau threefold $%
X_{3}^{\left( \mathbf{m},-\mathbf{p},\mathbf{2}\right) }$. Using the
realization eq(\ref{e}) and the splitting eq(\ref{sp}), the above local CY$%
^{3}$ reads as
\begin{equation}
\mathcal{O}(m_{0},..,m_{n})\oplus \mathcal{O}(-p_{0},..,-p_{n})\rightarrow
\left( \sum_{i=0}^{n}S_{i}^{2}\right) ,  \label{5}
\end{equation}%
with intersection matrix
\begin{equation}
\left[ S_{i}^{2}\right] .\left[ S_{j}^{2}\right] =-\hat{A}_{ij}.
\end{equation}%
The fibers $\mathcal{O}(\pm k_{0},..,\pm k_{n})$ carry charges under the various $%
U\left( 1\right) $ gauge symmetries of the individual 2-spheres $S_{i}^{2}$.
The total charge is given by eq(\ref{sp}). With these results at hand, we
are now in position to proceed forward and study the field theoretical
representation of the above class of local CY$^{3}$s by using the method of $%
2d$ $\mathcal{N}=2$ supersymmetric gauged linear sigma model.

\subsection{Supersymmetric field model}

Here we develop the study of type IIA geometry of $X_{3}^{\left( \mathbf{m},-%
\mathbf{p},\mathbf{2}\right) }$. To do so, we use known results on $2d$ $%
\mathcal{N}=2$ supersymmetric gauged linear sigma model formulation and take
advantage of our construction to also complete partial ones on type IIA
geometries based on standard affine models as well as non trivial fibrations.

To begin, recall that in the $2d$ $\mathcal{N}=2$ supersymmetric sigma model
framework, the field equations of motion of the auxiliary fields $D_{a}$ of
the gauge supermultiplets $V_{a}$
\begin{equation}
\frac{\delta \mathcal{S}_{2d}^{\mathcal{N}=2}}{\delta V_{a}}%
=\sum_{i}q_{i}^{a}\Phi _{i}^{\ast }\Phi _{i}-r^{a}=0  \label{v}
\end{equation}%
define the type IIA geometry. This method has been used in literature to
deal with K3 surfaces with ADE geometries. But here we would like to extend
this method to the case of the local geometry $X_{3}^{\left( \mathbf{m},-%
\mathbf{p},\mathbf{2}\right) }$. To that purpose, we shall proceed as
follows:\newline
(\textbf{1}) Study first the field realization on two special examples. This
analysis, which will be given in present subsection, allows to set up the
procedure. Then we give useful tools and illustrate the method.\newline
(\textbf{2}) Develop, as a next step, the general field theoretical $2d$ $%
\mathcal{N}=2$ supersymmetric gauged linear sigma model of eq(\ref{5}). This
is a more extensive and will be given in next section. Actually this is one
of the results of the present study.

\subsubsection{Type IIA model for $\mathcal{O}(m)\oplus \mathcal{O}(-m-2)\rightarrow S^{2}$}

\qquad To start note that for $m=0$, this local Calabi-Yau threefold
describes just the usual A$_{1}$ geometry on the complex line. The variety
has been studied extensively in literature; see \textrm{\cite{17}}\ for
instance. For non zero $m$ the situation is, as far as we know, new and its
supersymmetric linear sigma model can be obtained by considering a $U\left(
1\right) $ gauge field $V$ and four chiral superfields $\Phi _{i}$ with
gauge symmetry
\begin{equation}
\Phi _{i}\rightarrow \Phi _{i}^{\prime }=e^{iq_{i}\Lambda }\Phi _{i},\qquad
i=1,...,4,
\end{equation}%
with charges%
\begin{equation}
q_{i}=\left( 1,1,-2-m,m\right)
\end{equation}%
satisfying the CY condition $\sum_{i=1}^{4}q_{i}=0$. The gauge invariant
superfield action $\mathcal{S}_{2d}^{\mathcal{N}=2}$ of this model reads as
\begin{equation}
\mathcal{S}_{2d}^{\mathcal{N}=2}\sim \int d^{2}\sigma d^{4}\theta
\sum_{i=1}^{4}\Phi _{i}^{+}e^{q_{i}V}\Phi _{i}-r\int d^{2}\sigma d^{4}\theta
V
\end{equation}%
and the field equation of motion (\ref{v}) of the gauge field leads to
\begin{equation}
\left\vert \phi _{1}\right\vert ^{2}+\left\vert \phi _{2}\right\vert
^{2}-\left( 2+m\right) \left\vert \phi _{3}\right\vert ^{2}+m\left\vert \phi
_{4}\right\vert ^{2}=r.
\end{equation}%
This equation describes indeed $\mathcal{O}(m)\oplus \mathcal{O}%
(-m-2)\rightarrow S^{2}$. It has four special divisors $\phi _{i}=0$ while
the base 2-sphere corresponds to
\begin{equation}
\left\vert \phi _{1}\right\vert ^{2}+\left\vert \phi _{2}\right\vert ^{2}=r.
\end{equation}%
In case where $m$ is positive definite, this  geometry  can be also viewed as
describing the line bundle $\mathcal{O}(-m-2)$ over the weighted projective
space $\mathbb{WP}_{\left( 1,1,m\right) }^{2}$. Note that for $m=1$, one
gets the normal bundle of $\mathbb{P}^{2}$ and the local Calabi-Yau
threefold coincides with%
\begin{equation}
\mathcal{O}\left( -3\right) \rightarrow\mathbb{P}^{2}.
\end{equation}%
Note also that for $m=-1$, one has the resolved conifold
\begin{equation}
\mathcal{O}(-1)\oplus \mathcal{O}(-1)\rightarrow \mathbb{P}^{1}.
\end{equation}%
It is interesting to note here that resolved conifold can be realized from
normal bundle on $\mathbb{P}^{2}$ just by sending to infinity one of the edges of $%
\mathbb{P}^{2}$. Note finally that for $m=0,$ $-2$, one has the $A_{1}$
geometry fibered on the complex line $C$.

\subsubsection{Example: $\mathcal{O}(m_{1},m_{2})\oplus\mathcal{O}%
(-p_{1},-p_{2})\rightarrow \mathcal{C}_{2}$}

Following the same method, one can also build the type IIA sigma model for
this local CY$^{3}$ ($p_{i}=m_{i}+2$) based on two intersecting 2-spheres $%
S_{1}^{2}$ and $S_{2}^{2}$
\begin{equation}
\mathcal{C}_{2}=S_{1}^{2}+S_{2}^{2}
\end{equation}%
with intersection $S_{1}^{2}\cap S_{2}^{2}=\left\{ \text{ a point P}\right\}
$ modulo gauge transformations. The sigma model involves two $U\left(
1\right) $ gauge fields $V_{1}$, $V_{2}$ and five chiral superfields $\Phi
_{i}$. Denoting by $X_{1},$ $X_{2},$ $X_{3}$ the bosonic field components
parameterising the compact 2-cycle $\mathcal{C}_{2}$ and by $Y_{1},Y_{2}$
the complex variables parameterising $\mathcal{O}(-m_{1}-2,-m_{2}-2)$ and $%
\mathcal{O}(m_{1},m_{2})$ respectively, the two sigma model equations are
given by
\begin{eqnarray}
\left\vert X_{1}\right\vert ^{2}+\left\vert X_{2}\right\vert ^{2}-\left(
2+m_{1}\right) \left\vert Y_{1}\right\vert ^{2}+m_{1}\left\vert
Y_{2}\right\vert ^{2} &=&r_{1}  \notag \\
\left\vert X_{2}\right\vert ^{2}+\left\vert X_{3}\right\vert ^{2}-\left(
2+m_{2}\right) \left\vert Y_{1}\right\vert ^{2}+m_{2}\left\vert
Y_{2}\right\vert ^{2} &=&r_{2}.  \label{bt}
\end{eqnarray}%
In these equations, one recognizes two $SU\left( 2\right) /U\left( 1\right) $
relations describing the 2-spheres $S_{1}^{2}$ and $S_{2}^{2}$ associated
with taking $Y_{1}=Y_{2}=0;$ i.e
\begin{eqnarray}
\left\vert X_{1}\right\vert ^{2}+\left\vert X_{2}\right\vert ^{2} &=&r_{1}
\notag \\
\left\vert X_{2}\right\vert ^{2}+\left\vert X_{3}\right\vert ^{2} &=&r_{2}.
\label{s2}
\end{eqnarray}%
Notice that eqs(\ref{bt}) involve five complex (10 real) variables which are
not all of them free since they are constrained by two real constraint
relations (the D$_{1}$ and D$_{2}$ auxiliary field equations of motion) and $%
U\left( 1\right) \times U\left( 1\right) $ gauge symmetry
\begin{equation}
X_{1}\equiv X_{1}\exp \left( i\vartheta _{1}\right) ,\qquad X_{2}\equiv
X_{2}\exp \left( i\vartheta _{1}+i\vartheta _{2}\right) ,\qquad X_{3}\equiv
X_{3}\exp \left( i\vartheta _{2}\right) ,
\end{equation}%
where $\vartheta _{i}$ are the two gauge group parameters. At the end one is
left with $\left( 10-2-2\right) $ degrees of freedom which can be described
by three independent complex variables. Notice also that the spheres $%
S_{1}^{2}$ and $S_{2}^{2}$ intersect at
\begin{equation}
\left( X_{1},X_{2},X_{3}\right) =\left( r_{1},0,r_{2}\right) .  \label{fp}
\end{equation}%
$X_{2}=0$ is the fixed point under $U\left( 1\right) \times U\left( 1\right)
$ gauge symmetry (toric action) and up to a gauge transformation, the same
is valid for $X_{1}=r_{1}$ and $X_{3}=r_{2}$. Indeed if parameterising $%
X_{1}=\left\vert X_{1}\right\vert e^{-i\varphi }$ and $X_{3}=\left\vert
X_{3}\right\vert e^{-i\psi }$, one can ususally set $\varphi =\vartheta _{1}$
and $\psi =\vartheta _{2}$ by using $U\left( 1\right) \times U\left(
1\right) $ gauge invariance. Then setting $X_{2}=0$ in eqs(\ref{s2}), one
discovers eq(\ref{fp}). Notice finally that this construction generalizes
easily to the case of an open chain
\begin{equation}
\mathcal{C}_{n}=\sum_{i=1}^{n}S_{i}^{2},\qquad \mathcal{C}_{n}.\mathcal{C}%
_{n}=-2,
\end{equation}%
involving several intersecting 2-spheres $S_{i}^{2}$ with $\left[ S_{i}^{2}%
\right] .\left[ S_{i}^{2}\right] =-2$ and $\left[ S_{i}^{2}\right] .\left[
S_{i\pm 1}^{2}\right] =1$ otherwise zero. The sigma model field equations
describing this complex one dimension curve read as
\begin{equation}
\left\vert X_{i}\right\vert ^{2}+\left\vert X_{i+1}\right\vert
^{2}=r_{i},\qquad 1\leq i\leq n,  \label{bt3}
\end{equation}%
involving $\left( n+1\right) $ complex field variables $X_{i}$\ constrained
by $n$ complex constraint equations.
\begin{figure}[tbph]
\begin{center}
\hspace{0cm} \includegraphics[width=4cm]{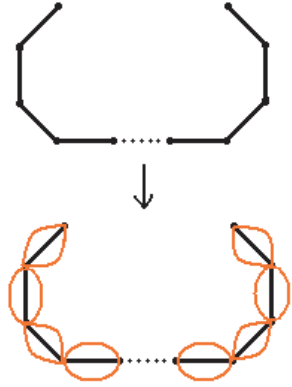}
\end{center}
\caption{{\protect\small {Open chain describing a typical compact base. It
involves several intersecting 2-spheres with one intersection point. Figure
in top involves straight lines and figure in bottom its fattening by
representing the circles above the 2-spheres.}}}
\end{figure}

\section{More on $2d$ $\mathcal{N}=2$ sigma model description}

\qquad So far we have considered special examples of type IIA realization of
local CY$^{3}$ as an introduction to the important case where the previous
(compact) open chain $\mathcal{C}_{n}$ gets closed by the adjunction of an
extra 2-sphere $S_{0}^{2}$,
\begin{equation}
\mathcal{C}_{n}\qquad \rightarrow \qquad \widehat{\mathcal{C}}_{n+1}
\end{equation}%
which we denote also as $\Delta _{n+1}$. In this section, we give the type
IIA description of these kinds of local CY$^{3}$s.

\begin{figure}[tbph]
\begin{center}
\hspace{0cm} \includegraphics[width=4cm]{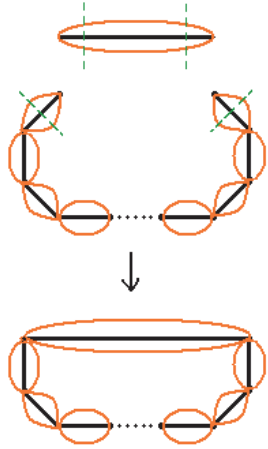}
\end{center}
\caption{{\protect\small {Toric graph of }$\Delta _{n+1}\ ${realized by
using affine }$\widehat{\mathit{A}}_{n}${\ Dynkin diagrams. It is obtained
by adding an extra 2-sphere in same manner as we do in building affine
Dynkin diagrams from ordinary ones.}}}
\end{figure}
To proceed, we start from the previous open chain $\mathcal{C}_{n}$ and add
an extra 2-sphere $S_{0}^{2}$,
\begin{equation}
\Delta _{n+1}=S_{0}^{2}+\mathcal{C}_{n},\qquad n>2,
\end{equation}%
with the following features,%
\begin{equation}
\left[ S_{0}^{2}\right] .\left[ S_{0}^{2}\right] =-2,\qquad \left[ S_{0}^{2}%
\right] .\left[ \mathcal{C}_{n}\right] =2.  \label{bt1}
\end{equation}%
In this case, we have
\begin{equation}
\left[ \Delta _{n+1}\right] .\left[ \Delta _{n+1}\right] =0,
\end{equation}%
as required by the homology property of the 2-torus. One of the solutions of
eq(\ref{bt}) is given by%
\begin{eqnarray}
\left[ S_{0}^{2}\right] .\left[ S_{1}^{2}\right] &=&1,\qquad \left[ S_{0}^{2}%
\right] .\left[ S_{n}^{2}\right] =1,  \notag \\
\left[ S_{0}^{2}\right] .\left[ S_{i}^{2}\right] &=&0,\qquad i\neq 1,\text{ }%
n.  \label{bt4}
\end{eqnarray}%
Other solutions are possible and are associated with Dynkin diagrams of
affine Kac-Moody algebras. To write down the explicit sigma model field
equations of the geometry $\mathcal{O}(\mathbf{m})\oplus \mathcal{O}(-%
\mathbf{p})\rightarrow \Delta _{n+1}$, we shall first write down the
equations for $\Delta _{n+1}$ and then give the general result.

\subsection{Sigma model eqs for $\Delta _{n+1}$}

A way to get the sigma model field equations for the 2-torus $\left[ \Delta _{n+1}%
\right] $, preserving the constraint equations; in particular the
dimensionality of $\Delta _{n+1}$ and $\left[ \Delta _{n+1}\right] \cdot %
\left[ \Delta _{n+1}\right] =0$, is to embed it in the $C^{n+3}$ with $%
\left( n+2\right) $ constraint complex relations resulting from $\left(
n+2\right) $ real equations of motion and $\left( n+2\right) $ abelian gauge
symmetries. The construction is done as follows: Start from eqs(\ref{bt3})
describing the open chain $\mathcal{C}_{n}$ together with the field equation
of motion
\begin{equation}
\left\vert Z_{0}\right\vert ^{2}+\left\vert Z_{1}\right\vert ^{2}=r_{0},
\label{bt5}
\end{equation}%
describing the extra 2-sphere $S_{0}^{2}$ with Kahler modulus $r_{0}$. The
meaning of the complex variables $Z_{0}$ and $Z_{1}$ will be specified
later. Then glue $S_{0}^{2}$ and $\mathcal{C}_{n}$ by implementing the
constraint relations (\ref{bt4}). A priori this could be achieved by setting
$Z_{1}=X_{1}$ and $Z_{0}=X_{n+1}$ so that eq(\ref{bt5}) becomes
\begin{equation}
\left\vert X_{n+1}\right\vert ^{2}+\left\vert X_{1}\right\vert ^{2}=r_{0},
\label{bt6}
\end{equation}%
In this way $S_{0}^{2}$ intersects once the 2-sphere $S_{1}^{2}$  defined by  $%
\left\vert X_{1}\right\vert ^{2}+\left\vert X_{2}\right\vert ^{2}=r_{1}$ as
well as $S_{n}^{2}$ with eq $\left\vert X_{n}\right\vert ^{2}+\left\vert
X_{n+1}\right\vert ^{2}=r_{n}$. But strictly speaking there is still a
problem although the resulting geometry looks having the topology of a
2-torus. This construction does not exactly work. The point is that by
combining eqs(\ref{bt3},\ref{bt6}), we cannot have the right dimension since
the $\left( n+1\right) $ complex variables $\left\{ X_{i},\text{ }1\leq
i\leq n+1\right\} $ are constrained by $\left( n+1\right) $ complex
constraint relations. As mentioned before, we have $\left( n+1\right) $ real
relations coming from the field equations of motion (\ref{bt3}-\ref{bt6})
and an equal number following from $U\left( 1\right) ^{n+1}$ gauge symmetry
acting on the field
\begin{equation}
X_{i}\qquad \rightarrow \qquad X_{i}e^{i\sum_{a=1}^{n+1}q_{i}^{a}\vartheta
_{a}}
\end{equation}%
with $i=1,...,n+1$ and the $\vartheta _{a}$'s being the $U\left( 1\right) $
group with charges $q_{i}^{a}=\left(
q_{1}^{a},q_{2}^{a},0,..,q_{n+1}^{a}\right) $ as follows,
\begin{eqnarray}
q_{i}^{1} &=&\left( 1,1,0,0,0,...,0,0,0\right)  \notag \\
q_{i}^{2} &=&\left( 0,1,1,0,0,...,0,0,0\right)  \notag \\
q_{i}^{3} &=&\left( 0,0,1,1,0,...,0,0,0\right)  \notag \\
&&... \\
q_{i}^{n} &=&\left( 0,0,0,0,0,...,1,1,0\right)  \notag \\
q_{i}^{n+1} &=&\left( 1,0,0,0,0,...,0,0,1\right) .  \notag
\end{eqnarray}%
This dimensionality problem can be solved in different, but a priori
equivalent, ways. Let us describe below the key idea behind these solutions.%
\newline
A natural way to do is to start from the complex two dimension ALE geometry
with blown up A$_{n}$ singularity and make an appropriate dimension
reduction down to one complex dimension. More precisely, start from
equations
\begin{equation}
\left\vert X_{j-1}\right\vert ^{2}-2\left\vert X_{j}\right\vert
^{2}+\left\vert X_{j+1}\right\vert ^{2}=r_{j},\qquad j=1,...,n  \label{g}
\end{equation}%
and add the following extra constraint relation reducing the dimensionality
by one
\begin{equation}
\left\vert X_{n}\right\vert ^{2}-2\left\vert X_{n+1}\right\vert
^{2}+\left\vert X_{0}\right\vert ^{2}=r_{0}.  \label{gg}
\end{equation}%
There is also an extra $U\left( 1\right) $ gauge invariance giving charges
to $\left( X_{n},X_{n+1},X_{0}\right) $. This picture involves $\left(
n+2\right) $ complex variables constrained by $\left( n+1\right) $
relations. The compact geometry is determined from eqs(\ref{g}-\ref{gg}) by
restricting to the compact divisors $X_{i}=0$. This gives%
\begin{eqnarray}
\left\vert X_{i-1}\right\vert ^{2}+\left\vert X_{i+1}\right\vert ^{2}
&=&r_{i}\qquad i=1,...,n  \notag \\
\left\vert X_{0}\right\vert ^{2}+\left\vert X_{n}\right\vert ^{2} &=&r_{0}.
\end{eqnarray}%
The second way is to use the correspondence%
\begin{equation}
\alpha _{i}\qquad \leftrightarrow \qquad S_{i}^{2}
\end{equation}
between roots of Lie algebras $\alpha _{i}$, which in present analysis are
given in terms of unit basis vectors of R$^{n+1}$ as,%
\begin{equation}
\alpha _{i}=e_{i}-e_{i+1}  \label{ei}
\end{equation}%
and the $S_{i}^{2}$ 2-sphere homology of ALE space with blowing up
singularities. To have the 2-torus, one should consider affine Kac-Moody
symmetries and use its correspondence between the imaginary root $\delta
=\alpha _{0}-\sum_{i}\alpha _{i}$
\begin{equation}
\delta \qquad \leftrightarrow \qquad T^{2}
\end{equation}%
Below, we shall develop an other way to do relying on the following
correspondence%
\begin{eqnarray}
e_{i}^{2}\qquad &\longleftrightarrow &\qquad \frac{1}{\rho }\left\vert
X_{i}\right\vert ^{2},  \notag \\
e_{i}\cdot e_{j}\qquad &\longleftrightarrow &\qquad \frac{1}{\rho }%
\left\vert Y_{{\small ij}}\right\vert ^{2},
\end{eqnarray}%
where $e_{i}$ is as in eqs(\ref{ei}). In the case where $\left\{
e_{i}\right\} $ are orthogonal; i.e $e_{i}.e_{j}=0$, there is no $Y_{{\small %
ij}}$ variable and this is interpreted as just a divisor equation. In this
correspondence, roots $\alpha _{i}$ are associated with cotangent bundle of $%
P^{1}$ since by computing $\alpha _{i}^{2}=e_{i}^{2}-2e_{i}.e_{{\small i+1}%
}+e_{{\small i+1}}^{2}$ and using above correspondence,%
\begin{equation}
\left\vert X_{i}\right\vert ^{2}-2\left\vert Y_{{\small i,i+1}}\right\vert
^{2}+\left\vert X_{i+1}\right\vert ^{2}=\rho
\end{equation}%
For $Y_{{\small i,i+1}}=0$, one recovers the usual 2-sphere. The link
between these ways initiated here will be developed in \textrm{\cite{88}}.

We start from eq(\ref{bt3}) and modify it as follows: (i) an extra complex
variable $X_{0}$ so that the new system involves the following complex
variables $\left\{ X_{0},X_{1},...,X_{n+1}\right\} $. The extra variable
charged under the $U_{1}\left( 1\right) $ abelian gauge symmetry,
\begin{equation}
X_{0}^{\prime }=X_{0}e^{-2i\vartheta _{0}},\qquad X_{1}^{\prime
}=X_{1}e^{i\vartheta _{0}},\qquad X_{2}^{\prime }=X_{2}e^{i\vartheta _{0}}.
\end{equation}%
(ii) an extra $U_{0}\left( 1\right) $ gauge symmetry acting as
\begin{equation}
X_{0}^{\prime }=X_{0}e^{i\vartheta _{0}},\qquad X_{1}^{\prime
}=X_{1}e^{i\vartheta _{0}}
\end{equation}%
and trivially on the remaining others. (iii) modify eq(\ref{bt3}) as
\begin{equation}
-2\left\vert X_{0}\right\vert ^{2}+\left\vert X_{1}\right\vert
^{2}+\left\vert X_{2}\right\vert ^{2}=r_{1},  \label{e}
\end{equation}%
whose compact part $X_{0}=0$ is just the 2-sphere $\left\vert
X_{1}\right\vert ^{2}+\left\vert X_{2}\right\vert ^{2}=r_{1}$ of eq(\ref{bt3}%
), and the remaining $\left( n-1\right) $ others unchanged
\begin{eqnarray}
\left\vert X_{2}\right\vert ^{2}+\left\vert X_{3}\right\vert ^{2} &=&r_{2}
\notag \\
... &=&...  \notag \\
\left\vert X_{n-1}\right\vert ^{2}+\left\vert X_{n}\right\vert ^{2}
&=&r_{n-1}, \\
\left\vert X_{n}\right\vert ^{2}+\left\vert X_{n+1}\right\vert ^{2} &=&r_{n}.
\notag
\end{eqnarray}%
(iv) Finally add moreover
\begin{equation}
\left\vert X_{n+1}\right\vert ^{2}+\left\vert X_{0}\right\vert ^{2}=r_{0}.
\label{f}
\end{equation}%
These relations involves $\left( n+2\right) $ complex variables $X_{i}$
subject to $\left( n+1\right) $ real constraint equations  and $\left( n+1\right) $
$U\left( 1\right) $ symmetries. They describe exactly the elliptic curve $%
\Delta _{n+1}$. Note that at the level eq(\ref{e}) the variable $X_{0}$
parameterises a complex space $C$, which in the language of toric graphs, is
represented by a half line. The relation (\ref{f}) describes then the
compactification of non compact complex space $C$ with variable $X_{0}$ to
the complex one projective space (real 2-sphere). \newline
In sigma model language, this corresponds to having $\left( n+2\right) $\
chiral superfields $\Phi _{i}$ with leading bosonic component fields,%
\begin{equation}
X_{0},\text{ \ }X_{1},\text{ \ }X_{2},\text{ ...\ ,..., }X_{n-1},\text{ \ }%
X_{n},\text{ \ }X_{n+1},
\end{equation}%
charged under $\left( n+1\right) $ Maxwell gauge superfields,%
\begin{equation}
V_{0},\text{ \ }V_{1},\text{ \ }V_{2},\text{ ...\ ,..., }V_{n-1},\text{ \ }%
V_{n},
\end{equation}%
with the $U^{n+1}\left( 1\right) $ charges $q_{i}^{a}=\left(
q_{0}^{a},q_{1}^{a},0,..,q_{n+1}^{a}\right) $ as follows,
\begin{equation*}
U_{0}\left( 1\right) :q_{i}^{0}=\left( 1,0,0,0,0,...,0,0,1\right)
\end{equation*}%
and%
\begin{eqnarray}
U_{1}\left( 1\right) &:&q_{i}^{1\text{ \ \ \ \ }}=\left(
-2,1,1,0,0,...,0,0,0\right)  \notag \\
U_{2}\left( 1\right) &:&q_{i}^{2\text{ \ \ \ \ }}=\left(
0,0,1,1,0,...,0,0,0\right)  \notag \\
&&... \\
U_{n-1}\left( 1\right) &:&q_{i}^{n-1}=\left( 0,0,0,0,0,...,1,1,0\right)
\notag \\
U_{n}\left( 1\right) &:&q_{i}^{n\text{ \ \ \ \ }}=\left(
1,0,0,0,0,...,0,1,1\right).  \notag
\end{eqnarray}%
Below, we use this contruction to build our class of local Calabi-Yau
threefold using the elliptic curve $\Delta _{n+1}$ as the compact part.

\subsection{Local 2-torus}

\qquad Implementing the fibers $\mathcal{O}(m)\oplus \mathcal{O}(-m)$ which,
in our present realization, take the form $\mathcal{O}(m_{0},..,m_{n})$ and $%
\mathcal{O}(-p_{0},..,-p_{n})$ and extending the construction of subsection
2.2, one can write down the sigma model relations. We have
\begin{eqnarray}
\left\vert X_{1}\right\vert ^{2}+\left\vert X_{2}\right\vert ^{2}-\left(
2+m_{1}\right) \left\vert X_{0}\right\vert ^{2}+m_{1}\left\vert
Y_{2}\right\vert ^{2} &=&r_{1}  \notag \\
\left\vert X_{2}\right\vert ^{2}+\left\vert X_{2}\right\vert ^{2}-\left(
2+m_{2}\right) \left\vert Y_{1}\right\vert ^{2}+m_{2}\left\vert
Y_{2}\right\vert ^{2} &=&r_{2}  \notag \\
&&...  \notag \\
\left\vert X_{n}\right\vert ^{2}+\left\vert X_{n+1}\right\vert ^{2}-\left(
2+m_{n}\right) \left\vert Y_{n}\right\vert ^{2}+m_{n}\left\vert
Y_{2}\right\vert ^{2} &=&r_{n}  \label{eq} \\
\left\vert X_{n+1}\right\vert ^{2}+\left\vert X_{0}\right\vert ^{2}-\left(
2+m_{0}\right) \left\vert Y_{1}\right\vert ^{2}+m_{0}\left\vert
Y_{2}\right\vert ^{2} &=&r_{0},  \notag
\end{eqnarray}%
where $Y_{1}$ and $Y_{2}$ are the fiber variables and carry non trivial
charges under $U^{n+1}\left( 1\right) $ gauge symmetries.

\section{Brane theory and 4d black holes in type II string }

\qquad In this section we consider type string IIA compactification on the
class of local Calabi-Yau threefolds constructed in previous sections
\begin{equation}
X_{3}^{\left( \mathbf{m,-m-2,2}\right) }=\mathcal{O}(\mathbf{m})\oplus
\mathcal{O}(-\mathbf{m-2})\rightarrow \Delta _{n+1}.
\end{equation}
Then, we develop a field theoretical method to study $4d$ large black holes
by using the 2d q-deformed quiver gauge theory living on $\Delta _{n+1}$.
Large black holes in  four dimensional space-time are generally obtained by using
configurations of type II or M-theory branes on cycles of the internal
manifolds. In type IIA framework, an interesting issue is given by\ BPS
configurations involving, amongst others, $N$ D4-branes wrapping non-compact
divisors of the local CY$^{3}$ giving rise to a dual of topological string.
Our construction follows more a less the same method used in \textrm{\cite%
{02}}; the difference comes mainly from the structure of the internal
manifold $X_{3}^{\left( \mathbf{m,-m-2,2}\right) }$ and the engineering of
the quiver gauge theory living on $\Delta _{n+1}$.

We first study the D-brane formulation of the BPS 4d black hole in the
framework of type IIA string compactification on local $\Delta _{n+1}$. Then
we study the reduction of $\mathcal{N}=4$ twisted topological theory on
4-cycles to 2d quiver gauge theory, represented by ADE Dynkin diagrams.

\subsection{Brane theory in $X_{3}^{\left( \mathbf{m,-m-2,2}\right) }$
background}

\qquad In type IIA string compactification on local Calabi-Yau threefolds $%
X_{3}^{\left( \mathbf{m,-m-2,2}\right) }$, the effective $4d$, $\mathcal{N}%
=2 $ supersymmetric theory has massive BPS particles coming from D-branes
wrapping cycles in $X_{3}^{\left( \mathbf{m,-m-2,2}\right) }$. Under some
assumptions, BPS states based on a special D-branes configuration may be
interpreted in terms of 4d space-time black holes. This configuration
involves D0-D4 and D2-D4 brane bound states but no D6 due to the reality of
the string coupling constant $g_{s}$. The D0-particles couple the RR type
IIA 1-form ${\large A}_{1}$ while the D2- and D4-branes couple to the RR
3-form ${\large C}_{3}$. Their respective charges $Q_{0}$, $Q_{2a}$ and $%
Q_{4}^{a}$ give the following expression of the macroscopic entropy of the
black hole $\cite{04,8,08}$
\begin{equation}
S_{BH}=\frac{1}{4}\sqrt{\frac{1}{6}C_{abc}Q_{4}^{a}Q_{4}^{b}Q_{4}^{c}\left(
Q_{0}-\frac{1}{2}C^{ab}Q_{2a}Q_{2b}\right) }  \label{fe}
\end{equation}%
with
\begin{eqnarray}
C_{abc} &=&\int_{CY3}\omega _{a}\wedge \omega _{b}\wedge \omega _{c},\qquad
\notag \\
C_{abc} &=&C_{abc}Q_{4}^{c},\qquad C^{ab}C_{bc}=\delta _{c}^{a}\text{ }.
\label{li}
\end{eqnarray}%
The above 4d black hole construction can be made more precise for our
present study. Here the local threefold $X_{3}^{\left( \mathbf{m,-m-2,2}%
\right) }$ is given by
\begin{equation}
\mathcal{O}(m_{0},..,m_{n})\oplus \mathcal{O}(-p_{0},..,-p_{n})\rightarrow
\left( \sum_{i=0}^{n}S_{i}^{2}\right) ,  \label{ci}
\end{equation}%
with $p_{i}=m_{i}+2$ . The 2-cycles basis $\left\{ [\mathcal{C}^{i}],\text{ }%
i=1,...,h^{1,1}(X)\right\} $ of $H_{2}(X,Z)$ is given by the compact
2-spheres $S_{i}^{2}$ with kahler modulus $r_{i}$ and the following
supersymmetric linear sigma field theoretical realization
\begin{equation}
\mathcal{C}^{i}:\left\vert X_{i}\right\vert ^{2}+\left\vert
X_{i+1}\right\vert ^{2}=r_{i},\qquad i=0,...,n,  \label{ta}
\end{equation}%
with the identification $S_{0}^{2}\equiv S_{n+1}^{2}$ and $n+1=h^{1,1}(X)$.
The components $[\mathcal{D}_{i}]$ of the dual basis of 4-cycles of $%
H_{4}(X,Z)$ is given by the non compact complex surfaces
\begin{equation}
\lbrack \mathcal{D}_{i}]=\mathcal{O}(-p_{0},..,-p_{n})\rightarrow \mathcal{C}%
^{i},\qquad i=0,...,n,  \label{ti}
\end{equation}%
with generic equations
\begin{equation}
\left\vert X_{i}\right\vert ^{2}+\left\vert X_{i+1}\right\vert ^{2}-\left(
2+m_{i}\right) \left\vert Z\right\vert ^{2}+m_{i}\left\vert Y_{2}\right\vert
^{2}=r_{i},\qquad i=0,...,n,  \label{on}
\end{equation}%
where $Z$ stands either for $X_{0}$\ or $Y_{1}$ as given in eqs(\ref{eq}).
These dual 2- and 4-cycles determine a basis for the $\left( n+1\right) $
abelian vector fields $B^{i}=B^{i}\left( t,\mathbf{r}\right) $ obtained by
integrating the RR 3-form ${\large C}_{3}$ on the 2-cycles $\mathcal{C}^{i}$
as shown below
\begin{equation}
B^{i}=\int_{\mathcal{C}^{i}}{\large C}_{3},\qquad \int_{\mathcal{C}^{i}}%
\mathbf{\omega }_{j}=\delta _{j}^{i}.  \label{po}
\end{equation}%
Under these $B^{i}$ abelian gauge fields, the D2-branes in the class $[%
\mathcal{C}]\in H_{2}(X,Z)$ and D4-branes in the class $[\mathcal{D}]\in
H_{4}(X,Z)$ are given by
\begin{equation}
\lbrack \mathcal{C}]=\sum_{i=0}^{n}Q_{2i}[\mathcal{C}^{i}],\qquad \lbrack
\mathcal{D}]=\sum_{i=0}^{n}Q_{4}^{i}[\mathcal{D}_{i}],  \label{ur}
\end{equation}%
and carry respectively $Q_{2i}$ ($Q_{2i}=M_{i}$) electric and $Q_{4}^{i}$ ($%
Q_{4}^{i}=N_{i}$) magnetic charges. We also have D0-brane charge $Q_{0}$
that couple the extra $U(1)$ vector field originating from RR 1-form.
D6-brane charges are turned off.

Following \textrm{\cite{02}}, the indexed degeneracy $\Omega \left(
Q_{0},Q_{2i},Q_{4}^{i}\right) $ of BPS particles in space-time with charges $%
Q_{0},$ $Q_{2i},$ $Q_{4}^{i}$ can be computed by counting BPS states in the
Yang-Mills theory on the D4-brane. This is computed by the supersymmetric
path integral of the four dimensional theory on $\mathcal{D}$ in the
topological sector of the Vafa-Witten maximally supersymmetric $\mathcal{N}%
=4 $ theory on $\mathcal{D}$ $\cite{02,vw,8}$,%
\begin{equation}
\mathcal{Z}_{Brane}=\int \left[ DA\right] \exp \left( -\frac{1}{2g_{s}}\int_{%
\mathcal{D}}TrF\wedge F-\frac{\theta }{g_{s}}\int_{\mathcal{D}}\mathbf{%
\omega }\wedge TrF\right).
\end{equation}%
Up to an appropriate gauge fixing, this relation can be written, by using
the chemical potentials $\varphi _{0}=\frac{4\pi }{g_{s}}$ and $\varphi _{1}=%
\frac{2\pi \theta }{g_{s}}$ for D0 and D2-branes respectively, as follows%
\begin{equation}
\mathcal{Z}_{Brane}\left[ N_{i},\varphi ^{0},\varphi ^{i}\right]
=\sum_{Q_{0},M_{i}}\Omega \left( Q_{0},M_{i},N_{i}\right) \exp \left(
-Q_{0}\varphi ^{0}-M_{i}\varphi ^{i}\right)
\end{equation}%
where we have used%
\begin{equation}
Q_{0}=\frac{1}{8\pi ^{2}}\int_{\mathcal{D}}\mathrm{Tr}\left( F\wedge
F\right) ,\qquad M_{i}=\frac{1}{2\pi }\int_{\mathcal{D}}\mathrm{Tr}F\wedge
\mathbf{\omega }_{i}.
\end{equation}%
The above relation may be expanded in series of $e^{-g_{s}}$ due to S
duality of underlying $\mathcal{N}=4$ theory that relates strong and weak
coupling expansions \cite{8}. Recall that the world-volume gauge theory on
the $N$ D4-branes is the $\mathcal{N}=4$ topological $U(N)$ YM on $\mathcal{D%
}$. Turning on chemical potentials for D0-brane and D2-brane correspond to
introducing the observables
\begin{equation}
\mathcal{S}=\frac{1}{2g_{s}}\int_{\mathcal{D}}\mathrm{Tr}F\wedge F+\frac{%
\theta }{g_{s}}\int_{\mathcal{D}}\mathbf{\omega }\wedge \mathrm{Tr}F
\label{top}
\end{equation}%
where $\mathbf{\omega }$ is the unit volume form of $\Delta _{n+1}$. The
topological theory (\ref{top}) is invariant under turning on the massive
deformation
\begin{equation}
\delta \mathcal{S}=\sum_{i}\frac{p_{i}}{2}\int \Phi _{i}^{2}
\end{equation}%
which simplifies the theory. By using further deformation which correspond
to a U(1) rotation on the fiber, the theory localizes to modes which are
invariant under the U(1) and effectively reduces the 4d theory to a gauge
theory on $\Delta _{n+1}$.

\subsection{2d quiver gauge theories on $\Delta _{n+1}$}

\qquad Note first that from four dimensional space-time view, the wrapped $N$
D-branes on $\mathcal{D}=\mathcal{O}(-p)\rightarrow \mathcal{C}$ describe a
point-particle with dynamics governed by $4d$ $\mathcal{N}=2$ supergravity
coupled to $U\left( N\right) $ super-Yang Mills. On the D4-branes live a $%
\left( 4+1\right) $ space-time $\mathcal{N}=4$ $U\left( N\right) $
supersymmetric gauge theory and on its reduced topological sector one has a
4d $\mathcal{N}=4$ topological theory twisted by massive deformations.

In our present study, the 2-cycle $\mathcal{C}$ is represented by the closed
chain $\Delta _{n+1}$ with multi- toric actions and the line $\mathcal{O}(-p)$ is a
non trivial fiber capturing charges under these abelian symmetries. It will
be denoted as $\mathcal{O}(-\mathbf{p})$,
\begin{equation}
\mathbf{p}=\left( p_{1},...,p_{n}\right) .  \label{ic}
\end{equation}%
As a consequence of the topology of $\Delta _{n+1}$ which is given by $%
\left( n+1\right) $ intersecting 2-spheres, the previous $U\left( N\right) $
gauge invariance gets broken down to%
\begin{equation}
U\left( N\right) \qquad \rightarrow \qquad U(N_{0})\times U(N_{1})...\times
U\left( N_{n}\right) ,  \label{ob}
\end{equation}%
with the following condition on the group ranks
\begin{equation}
N=\sum_{i=0}^{n}N_{i}.  \label{tn}
\end{equation}%
This symmetry breaking phenomenon requires non zero Kahler moduli (\ref{r})
of the various 2-spheres of the base $\Delta _{n+1}$,
\begin{equation}
r_{i}\neq 0,\qquad i=0,1,...,n.  \label{gs}
\end{equation}%
Viewed from 4d space-time, the effective theory of type IIA string low
energy limit on local CY$^{3}$ with D-branes wrapping cycles is given by a $%
4d$ $\mathcal{N}=2$ supergravity coupled to $4d$ $\mathcal{N}=2$ quiver
gauge theory with gauge group $G=U(N_{0})\times ...\times U\left(
N_{n}\right) $. This is the general picture of the string low energy
effective field approximation.\newline
By requiring $4d$ space-time $\mathcal{N}=2$\ super-conformal invariance,
the vanishing conditions $\beta _{i}=0$ of one loop beta functions on the
gauge group factors $U(N_{i})$ put a strong constraint  on the ranks
$N_{i}$ of these $U(N_{i})$'s. These conditions have been approached for
different purposes; in particular in the context of geometric engineering of
$4d$ space-time $\mathcal{N}=2$ superconformal theories embedded in type IIA
string theory  on local CY$^{3}$s with blown up affine ADE singularities. There,
these conditions take the remarkable form
\begin{equation}
\sum_{i=0}^{n}\widehat{K}_{ij}N_{j}=0,  \label{sra}
\end{equation}%
and  are  solved by taking $N_{i}$ ranks as $N_{i}=s_{i}M$, where the $s_{i}$%
's\ are the Dynkin weights introduced earlier. In the case of affine $\hat{A}%
_{n}$ model, the $s_{i}$'s are equal to unity and so the $U\left( N\right) $
gauge group gets reduced, in the superconformal case, to 
\begin{equation}
U\left( N\right) \qquad \rightarrow \qquad G_{scft}=U(M)^{n+1}  \label{at}
\end{equation}%
with
\begin{equation}
N=\left( n+1\right) M.  \label{mu}
\end{equation}%
On the 4-cycle $\mathcal{D}$ of the local Calabi-Yau threefold, the theory
is a $\mathcal{N}=4$ topologically twisted gauge theory; but using the
result of \textrm{\cite{02}}, this theory can be simplified by integrating
gauge field configuration on fiber $\mathcal{O}(-\mathbf{p})$ and fermionic degrees
freedom to end with a 2d bosonic quiver gauge theory on $\Delta _{n+1}$.
This theory has $\dprod\nolimits_{i}U(N_{i})$ as a gauge symmetry group and
involves:\newline
(\textbf{i}) Gauge fields $A_{i}$, $i=0,1,..,n$, for each gauge group factor
$U(N_{i})$ with field strength $F_{i}=dA_{i}+A_{i}\wedge A_{i}$,%
\begin{equation}
F_{i}=\sum_{a_{i}=1}^{N_{i}}H_{a_{i}}F_{i}^{a_{i}}+\sum_{\beta _{i}\text{ of
}U(N_{i})}E_{\beta _{i}}^{+}F_{i}^{-\beta _{i}}+E_{\beta
_{i}}^{-}F_{i}^{+\beta _{i}},  \label{x}
\end{equation}%
where $\left\{ H_{a_{i}},E_{\beta _{a_{i}}}^{\pm }\right\} $ is the Cartan
basis of $U(N_{i})$ and where $\beta _{i}$ are the positive roots\footnote{%
We have used the Greek letter $\beta $ to refer to the roots of the of the
gauge group $U\left( N\right) $. Positive roots of the $U\left( N_{i}\right)
$\ are denoted by $\beta _{i}$ and should not confused with simple roots $%
\alpha _{i}$ used in the intersection matrix $\left( K_{ij}=\alpha
_{i}.\alpha _{j}\right) $ of the 2-cycles of the base $\Delta _{n}$ of local
CY3.} of the $i$-th gauge group factor. The above relation contains as a
closed subset the usual $U^{N_{i}}\left( 1\right) $ abelian part $%
dA_{i}=\sum_{a_{i}=1}^{N_{i}}H_{a_{i}}dA_{i}^{a_{i}}$ where the $H_{a_{i}}$%
's\ are the commuting Cartan generator. \newline
(\textbf{ii}) 2d scalars $\Phi _{i}=\Phi _{i}\left( z,\overline{z}\right) $
in the adjoint for each factor $U(N_{i})$ having a similar expansion as in (%
\ref{x}); but which we reduce to its $U^{N_{i}}\left( 1\right) $ diagonal
form
\begin{equation}
\Phi _{i}=\sum_{a_{i}=1}^{N_{i}}H_{a_{i}}\Phi _{i}^{a_{i}},\qquad \Phi
_{i}^{\pm \beta _{i}}=0,\qquad i=0,...,n.
\end{equation}%
These fields are obtained by integration of the 4d gauge field strengths $%
F_{i}\left( z,\overline{z},y,\overline{y}\right) $ on the fiber $\mathcal{O}%
\left( -\mathbf{p}\right) $
\begin{equation}
\Phi _{i}\left( z,\overline{z}\right) =\int_{\mathcal{O}\left( -\mathbf{p}%
\right) }d^{2}y\text{ }F_{i}\left( z,\overline{z},y,\overline{y}\right)
\end{equation}%
which, as usual, can be put as
\begin{equation*}
\Phi _{i}\left( z,\overline{z}\right) =\doint\nolimits_{S_{i}^{1},\text{ }%
\left\vert y\right\vert \rightarrow \infty }A_{i}\left( z,\overline{z},y,%
\overline{y}\right) ,\qquad z\in S_{i}^{2}
\end{equation*}%
where the loop $S_{i}^{1}$ can be thought of as a circle at the infinity ($%
\left\vert y\right\vert \rightarrow \infty $) of the non compact fiber $%
\mathcal{O}\left( -p_{i}\right) \sim C$ parameterized by the complex variable $y$.%
\newline
(\textbf{iii}) 2d matter fields $\Phi _{ij}$ in the bi-fundamentals of the
quiver gauge group living on the intersection of $\Delta _{n+1}$ patches
with, in general, a leg on $S_{i}^{2}$ and the other on $S_{j}^{2}$.\newline
In the language of the representations of the gauge symmetry $U\left(
N_{i}\right) \times U\left( N_{j}\right) $, these fields belongs to $\left(
N_{i},\overline{N}_{j}\right) $ and describe the link between the gauge
theory factors living on the irreducible 2-cycles making $\Delta _{n+1}$.

In the language of topological string theory using caps, annuli and
topological vertex \cite{4}, these bi-fundamentals can be implemented in the
topological partition function thought insertion operators type
\begin{equation}
S_{\mathcal{R}_{i}\overline{\mathcal{R}}_{i+1}}
\end{equation}%
using sums over representations $\mathcal{R}_{i}\overline{\mathcal{R}}_{i+1}$
of the gauge invariance $U\left( N_{i}\right) \times U\left( N_{i+1}\right) $.
 This construction has been studied recently in \textrm{\cite{8}} for
particular classes of local CY$^{3}$ such as $\mathcal{O}\left( -3\right) \rightarrow
\mathbb{P}^{2}$ and $\mathcal{O}\left( -2,-2\right) \rightarrow \mathbb{P}^{1}\times\mathbb{P}^{1}$. We will not
develop this issue here\footnote{%
As argued in \cite{8}, the matter fields localized at the intersection point
$P_{i}$ of the 2-spheres $S_{i}^{2}$ and $S_{i+1}^{2}$ corresponds to
inserting the operator $\mathcal{V}=\sum_{\mathcal{R}}Tr_{\mathcal{R}%
}V_{i}^{-1}Tr_{\mathcal{R}}V_{\left( i+1\right) }$ with $V_{i}=\exp \left(
i\Phi _{\left( i\right) }-i\doint A_{\left( i\right) }\right) $ and $%
V_{\left( i+1\right) }=e^{i\Phi _{\left( i+1\right) }}$ where the integral
contour is a small loop around $P_{i}$.}. Below we shall combine however
field theoretical analysis and representation group theoretical method to
deal with bi-fundamentals.

\subsubsection{Derivation of 2d quiver gauge field action}

\qquad Here we construct the 2d field action $\mathcal{S}_{\Delta _{n+1}}$
describing the localization of the topological gauge theory of the BPS D4-,
D2-, D0-brane configurations on the non compact divisor $\left[ \mathcal{D}%
\right] =\mathcal{O}(-\mathbf{p})\rightarrow \Delta _{n+1}$ of the local CY$%
^{3}$. This action can be obtained by following the same method as done for
the case $\mathcal{O}\left( -p\right) \rightarrow \Sigma _{g}$. \newline
One starts from eq(\ref{top}) describing the gauge theory on the $N$
D4-branes wrapping $\mathcal{D}$ with D0-D4 and D2-D4 bound states
\begin{equation}
\mathcal{S}_{4d}=\frac{1}{2g_{s}}\int_{\left[ \mathcal{D}\right] }\mathrm{Tr}%
\left( F\wedge F\right) +\frac{\theta }{g_{s}}\int_{\left[ \mathcal{D}\right]
}\mathrm{Tr}F\wedge \mathbf{\omega }.  \label{dda}
\end{equation}%
In this equation, the parameters $g_{s}$ and $\theta $ are related to the
chemical potentials $\varphi _{0}$ and $\varphi _{1}$ for D0 and D2-branes
respectively as $\varphi _{0}=\frac{4\pi ^{2}}{g_{s}}$ and $\varphi _{1}=%
\frac{2\pi \theta }{g_{s}}$. The field $F$ is the 4d $U\left( N\right) $
gauge field strength $F=dA+A\wedge A$. It is a hermitian 2-form with gauge
connexion $A$. The field 1-form $A$ reads in the local coordinates $\left\{
z,\overline{z},y,\overline{y}\right\} $, with $\left( z,\overline{z}\right) $
parameterising the base $\Delta _{n+1}$ and $\left( y,\overline{y}\right) $
for the fiber, as follows
\begin{eqnarray}
A &=&A_{z}dz+A_{y}dy+\overline{A}_{\bar{z}}d\overline{z}+\overline{A}_{\bar{y%
}}d\overline{y},\qquad  \notag \\
A_{\mu } &=&A_{\mu }\left( z,\overline{z},y,\overline{y}\right) ,\qquad \mu
=z,y,\overline{z},\overline{y}.
\end{eqnarray}%
Moreover, like $A$, the 2-form field $F$ is valued in the Lie algebra of $%
U\left( N\right) $ gauge symmetry and so can be expanded as
\begin{equation}
F=\sum_{a=1}^{N}H_{a}F^{a}+\sum_{\text{positive roots }\beta \text{ of }%
U(N)}E_{\beta }^{+}F^{-\beta }+E_{\beta }^{-}F^{+\beta },
\end{equation}%
where $\left\{ H_{a_{i}},E_{\beta _{i}}^{\pm }\right\} $ is the Cartan basis
of $U(N_{i})$. The above relation contains as a closed subset the usual $%
U^{N}\left( 1\right) $ abelian part
\begin{equation}
dA=\sum_{a=1}^{N}H_{a}dA^{a},\qquad \left[ H_{a},H_{n}\right] =0,  \label{ab}
\end{equation}%
which plays a crucial role in the computation of Wilson loops. In eq(\ref%
{dda}), $\mathbf{\omega }$ is the 2-form on the compact cycle $\Delta _{n+1}$
on which the D2-branes lives and is normalized as%
\begin{equation}
\int_{\Delta _{n+1}}\mathbf{\omega }=1.
\end{equation}%
On the other hand, using eq(\ref{el}-\ref{ef}), we can put the right hand
side of the above relation\footnote{%
We have used the formula $\int_{A\cup B}K=\int_{A}K+\int_{B}K-\int_{A\cap
B}K $ for Kahler modulus of two intersecting surfaces.} as follows
\begin{equation}
\int_{\Delta _{n}}\mathbf{\omega }=\sum_{i}\left( \int_{S_{i}^{2}}\mathbf{%
\omega }\right) -\frac{1}{2}\sum_{i\neq j}\left( \int_{S_{i}^{2}\cap
S_{j}^{2}}\mathbf{\omega }\right) .  \label{ad}
\end{equation}%
Note that since $\int_{S_{i}^{2}\cap S_{j}^{2}}\mathbf{\omega }$ vanishes
for the intersections $S_{i}^{2}\cap S_{j}^{2}$ which are given by a set of
separated points, we can simplify this expression further into
\begin{equation}
\int_{\Delta _{n}}\mathbf{\omega }=\frac{1}{r}\sum_{i=0}^{n}r_{i}=1,\qquad
r\neq 0.
\end{equation}%
This shows that on $\Delta _{n+1}=\sum_{i=0}^{n}\mathcal{C}^{i}$, the Kahler
form splits as $\mathbf{\omega }=\frac{1}{r}\sum_{i=0}^{n}r_{i}\mathbf{%
\omega }_{i}$ with $\int_{\mathcal{C}^{i}}\mathbf{\omega }_{j}=\delta
_{j}^{i}$.\newline
The next step is to perform integration on the fiber variables $y$ and $%
\overline{y}$.  The topological theory (\ref{dda}) localizes to modes which
are invariant under $U^{n+1}(1)$ symmetries and effectively reduces to a
gauge theory on the base $\Delta _{n+1}$. Let us give details by working out
explicitly these steps: (i) First, we have
\begin{eqnarray}
\int_{\left[ \mathcal{D}_{4}\right] }\mathrm{Tr}\left( F\wedge F\right)
&=&\int_{z\in \Delta _{n+1}}\mathrm{Tr}\left( \int_{\mathcal{O}(-\mathbf{p})%
\text{ }\rightarrow \text{ }z}F\wedge F\right)  \notag \\
&=&2\int_{\Delta _{n}}\mathrm{Tr}\left( \Phi F\right)
\end{eqnarray}%
where we have set
\begin{equation}
\Phi \left( z,\overline{z}\right) =\int_{\mathcal{O}(-\mathbf{p})\text{ }%
\rightarrow \text{ }z}d^{2}yF\left( y,\overline{y},z,\overline{z}\right)
\end{equation}%
and where we have restricted $F$ to its values in the $U^{N}\left( 1\right) $
abelian subsalgebra eq(\ref{ab}) in order to put $\Phi \left( z,\overline{z}%
\right) $ in the Wilsonian form
\begin{equation}
\Phi \left( z,\overline{z}\right) =\doint\nolimits_{S_{\left\vert
y\right\vert }^{1}\text{ }\rightarrow \text{ }z,\text{ }\left\vert
y\right\vert \rightarrow \infty }A\left( y,\overline{y},z,\overline{z}%
\right) ,\qquad z\in \Delta _{n+1}.
\end{equation}%
The same thing can be done for the second term of eq(\ref{dda}). We find
\begin{equation}
\int_{\left[ \mathcal{D}\right] }\mathbf{\omega }\wedge \left( \mathrm{Tr}%
F\right) =\int_{\Delta _{n+1}}\mathbf{\omega }\mathrm{Tr}\left( \Phi \right).
\end{equation}%
The 4d action (\ref{dda}) reduces then to the following 2d one%
\begin{equation}
\mathcal{S}_{2d}=\frac{1}{g_{s}}\int_{\Delta _{n+1}}\mathrm{Tr}\left( \Phi
F\right) +\frac{\theta }{g_{s}}\int_{\Delta _{n+1}}\mathbf{\omega }\mathrm{Tr%
}\Phi .  \label{y0}
\end{equation}%
Now using the fact that $\Delta _{n+1}=\sum_{i=0}^{n}\mathbb{P}_{i}^{1}$ combined
with eq(\ref{ad}), we see that, depending on the patches of $\Delta _{n+1}$
where the Wilson field $\Phi $ is sitting, we get either adjoint 2d scalars $%
\Phi _{i}$ or bi-fundamentals $\Phi _{ij}$ as shown below%
\begin{eqnarray}
\Phi \left( z\right) &\equiv &\Phi _{i}\left( z\right) ,\qquad z\in
\mathbb{P}_{i}^{1},  \notag \\
\Phi \left( z\right) &\equiv &\Phi _{ij}\left( z\right) ,\qquad z\in
\mathbb{P}_{i}^{1}\cap \mathbb{P}_{j}^{1}.  \label{y1}
\end{eqnarray}%
Note that the $\Phi _{i}$ fields are valued in $U^{N_{i}}\left( 1\right) $
maximal abelian group. They parameterise the maximal $T^{N_{i}}$ torii of
the Lie group $U\left( N_{i}\right) $. So they should be compact and undergo
periodicity conditions. This means that the linear expansion
\begin{equation}
\Phi _{i}=\sum_{a=1}^{N}\Phi _{a_{i}}H_{a_{i}},\qquad \Phi _{a_{i}}\sim
Tr\left( H_{a_{i}}\Phi _{i}\right) ,  \label{be}
\end{equation}%
should be understood as
\begin{equation}
U_{i}=\exp i\Phi _{i},\qquad i=0,...,n,
\end{equation}%
and so the 2d field components $\Phi _{a_{i}}$ are constrained as,%
\begin{equation}
\Phi _{a_{i}}\equiv \Phi _{a_{i}}+2\pi m_{a_{i}},\qquad m_{a_{i}}\in Z,
\label{pe}
\end{equation}%
 leaving $U_{i}$ invariant.\newline
Now substituting eq(\ref{y1}) in the relation (\ref{y0}), we obtain after
implementing the hermiticity condition $\Phi =\frac{1}{2}\left( \Phi +\Phi
^{\ast }\right) $ and $F=\frac{1}{2}\left( F+F^{\ast }\right) $, the
following%
\begin{equation}
\int_{\Delta _{n}}\mathrm{Tr}\left( \Phi F\right) =\sum_{i}\int_{P_{i}^{1}}%
\mathrm{Tr}\left( \Phi _{i}F_{i}\right) -\frac{1}{8}\sum_{i\neq
j}\int_{P_{i}^{1}\cap P_{j}^{1}}\mathrm{Tr}\left( \Phi _{ij}F_{ji}+\mathrm{cc%
}\right)
\end{equation}%
where we have set $F_{ji}=F_{ij}^{\ast }$ and where we have disregarded the
terms $\Phi _{ij}F_{ij}$ transforming in the $\left( N_{i}^{\otimes 2},%
\overline{N}_{j}^{\otimes 2}\right) $ representation of $U(N_{i})\times
U\left( N_{j}\right) $. These terms do not preserve the abelian subsymmetry $%
U^{n+1}\left( 1\right) $ of the quiver gauge group $U(N_{0})\times ...\times
U\left( N_{n}\right) $. These 2d field configurations have a group
theoretical interpretation. They correspond to splitting the adjoint
representation of $U\left( N\right) $ with $N=N_{0}+...+N_{n}$ in terms of
representation of $U(N_{0})\times ...\times U\left( N_{n}\right) $%
\begin{equation}
\mathrm{Adj}U\left( N\right) =\sum_{i=0}^{n}\mathrm{Adj}U\left( N_{i}\right)
\oplus \sum_{i\neq j}\left( N_{i},\overline{N}_{j}\right).
\end{equation}%
The terms of the first sum are associated with $\Phi _{i}\left( z\right) $
while the other are associated with $\Phi _{ij}$. Obviously since in present
case only $\mathbb{P}_{i}^{1}\cap \mathbb{P}_{i\pm 1}^{1}$ which are non trivial, there are no
bi-fundamentals $\Phi _{ij}$ for $j\neq i\pm 1$.\newline
For the term $\int_{\Delta _{n}}\mathbf{\omega }\mathrm{Tr}\left( \Phi
\right) $ (\ref{y0}), we get%
\begin{equation}
\int_{\Delta _{n+1}}\mathrm{Tr}\left( \Phi \right) \mathbf{\omega }%
=\sum_{i=0}^{n}\frac{r_{i}}{r}\int_{P_{i}^{1}}\mathrm{Tr}\left( \Phi
_{i}\right) \mathbf{\omega }_{i}.
\end{equation}%
Let us first discuss these configurations separately and then give the
general result.\newline
\textbf{Adjoint 2d scalars}\newline
Putting eq(\ref{f}) back into (\ref{dda}) and focusing on the patches $%
\mathcal{P}_{i}^{1}$ by substituting $\Phi _{diag}=\sum_{i}\Phi _{i}\left( z\right) $,
we get the diagonal part of the topological 2d quiver gauge field action
\begin{equation}
\mathcal{S}_{diag}=\sum_{i=0}^{n}\mathcal{S}_{i}
\end{equation}%
with%
\begin{eqnarray}
\mathcal{S}_{i} &=&\frac{1}{g_{s}}\int_{\mathbb{P}_{i}^{1}}\mathrm{Tr}\left(
\Phi _{i}F\right) +\frac{\theta }{g_{s}}\frac{r_{i}}{r}\int_{\mathbb{P}%
_{i}^{1}}\mathrm{Tr}\Phi _{i}  \notag \\
&&+\frac{p_{i}}{2g_{s}}\int_{S_{i}^{2}}\mathrm{Tr}\Phi _{i}^{2},
\end{eqnarray}%
and where we have added the topologically invariant point-like observables $%
\mathrm{Tr}\Phi _{i}^{2}$ at the points $z\in \mathcal{P}_{i}^{1}$. Upon integrating
out fermions and adjoint scalars using $\Phi _{i}=\frac{-F_{i}-\theta _{i}}{%
p_{i}}$ and following \cite{02}, this topologically twisted theory is
equivalent to the bosonic 2d Yang-Mills theory
\begin{equation}
\mathcal{S}_{i}=-\int_{\mathbb{P}_{i}^{1}}\frac{1}{2g_{i}^{2}}\mathrm{Tr}%
\left( F_{i}^{2}\right) -\frac{\theta _{i}^{YM}}{g_{i}^{2}}\int_{\mathbb{P}%
_{i}^{1}}\mathrm{Tr}F_{i}
\end{equation}%
with Yang-Mills gauge coupling constants $g_{{\small YMi}}\equiv g_{i}$ and $%
\theta _{i}^{{\small YM}}\equiv \theta _{i}$ terms given by,%
\begin{equation}
g_{i}^{2}=p_{i}g_{s}\frac{r_{i}}{r},\qquad \theta _{i}^{YM}=\theta \frac{%
r_{i}}{r}.  \label{ts}
\end{equation}%
Here the $r_{i}$'s are the Kahler parameters of the 2-sphere constituting $%
\Delta _{n+1}$. Note that these gauge coupling constants and $\theta _{i}$'s
are not completely independent and are related amongst others as
\begin{equation}
\frac{\left( n+1\right) r}{g_{s}}=\sum_{i=0}^{n}\frac{p_{i}r_{i}}{g_{i}^{2}}%
,\qquad \sum_{i=0}^{n}\theta _{i}^{YM}=\theta .
\end{equation}%
The first equation should be compared with the standard relation $\frac{1}{%
g_{s}}=\sum_{i=0}^{n}g_{i}^{-2}$ appearing in the geometric engineering of
quiver gauge theories.\newline
\textbf{2d bi-fundamentals}\newline
To get the field action describing the contribution of bi-fundamentals, it
is interesting to proceed in steps as follows:\newline
Start from the topological field action on the 4-cycle $\left[ \mathcal{D}%
_{4}\right] $,
\begin{equation}
\mathcal{S}_{4d}=\frac{1}{2g_{s}}\int_{\left[ \mathcal{D}_{4}\right] }%
\mathrm{Tr}\left( F\wedge F\right) +\frac{\theta }{g_{s}}\int_{\left[
\mathcal{D}_{4}\right] }\mathrm{Tr}F\wedge \mathbf{\omega }
\end{equation}%
and think about $F$ as a field strength valued in the maximal non abelian
gauge group $U\left( N_{0}+...+N_{n}\right) $. Then expand the real field F
as
\begin{equation}
F=\sum_{i=0}^{n}F_{i}+\sum_{i<j}\left( F_{ij}+F_{ji}\right) ,\qquad F_{ji}=%
\overline{\left( F_{ij}\right) }  \label{ex}
\end{equation}%
with $F_{ji}=\overline{\left( F_{ij}\right) }$. The $F_{i}$'s are the real
field strengths of the gauge fields $A_{i}$ valued in the adjoints of $%
U(N_{i})$ factors with generators $\left\{ T_{a_{i}}\right\} $
\begin{equation}
F_{i}=\sum_{a_{i}=1}^{N_{i}^{2}}F_{i}^{a_{i}}T_{a_{i}},\qquad
i=1,..,n,\qquad F_{i}=\overline{\left( F_{i}\right) }.
\end{equation}%
The $F_{ij}$'s are the field strengths of the gauge fields $A_{ij}$ valued
in the Lie algebra associated to the coset
\begin{equation}
\frac{U\left( N_{0}+...+N_{n}\right) }{U(N_{0})\times ...\times U\left(
N_{n}\right) }.
\end{equation}%
Obviously the group $U\left( N_{0}+...+N_{n}\right) $ is not a full gauge
invariance of the $\mathcal{N}=4$ topological gauge theory since the gauge
fields $A_{ij}$ part get non zero masses \textrm{m}$_{ij}\sim \left(
r_{i}-r_{j}\right) $ after breaking $U\left( N_{0}+...+N_{n}\right) $ down
to $U(N_{0})\times ...\times U\left( N_{n}\right) $. \newline
The next step is to use the same trick as before by integrating partially
over the variables of the fiber $\mathcal{O}(-p_{0},..,-p_{n})$. We get%
\begin{eqnarray}
\mathcal{S}_{4d} &=&\frac{1}{g_{s}}\sum_{i}\int_{\mathbb{P}_{i}^{1}}\mathrm{%
Tr}\left( \Phi _{i}\wedge F\right) +\frac{\theta }{g_{s}}\sum_{i}\int_{%
\mathbb{P}_{i}^{1}}\mathrm{Tr}\Phi _{i}\wedge \mathbf{\omega }  \notag \\
&&-\frac{1}{g_{s}}\sum_{i\neq j}\int_{\mathbb{P}_{i}^{1}\cap \mathbb{P}%
_{j}^{1}}\mathrm{Tr}\left( \Phi _{ij}\wedge F\right) -\frac{\theta }{g_{s}}%
\sum_{i\neq j}\int_{\mathbb{P}_{i}^{1}\cap \mathbb{P}_{j}^{1}}\mathrm{Tr}%
\Phi _{ij}\wedge \mathbf{\omega }.  \label{sr}
\end{eqnarray}%
Now using the expansion (\ref{ex}) and the property
\begin{equation}
\mathrm{Tr}\left( T_{a_{i}}T_{b_{j}}\right) \sim \delta _{ab}\delta _{ij}
\end{equation}%
we have%
\begin{eqnarray}
\mathrm{Tr}\left( \Phi _{i}\wedge F\right) &=&\mathrm{Tr}\left( \Phi
_{i}\wedge F_{i}\right) ,\quad  \notag \\
\mathrm{Tr}\left( \Phi _{ij}\wedge F\right) &=&\mathrm{Tr}\left( \Phi
_{ij}\wedge F_{ji}\right) ,\quad \\
\mathrm{Tr}\left( \Phi _{ij}\right) &=&0.  \notag
\end{eqnarray}
Then we can bring eq(\ref{sr}) into the following reduced form
\begin{eqnarray}
\mathcal{S}_{2d} &=&\frac{1}{g_{s}}\sum_{i}\int_{\mathbb{P}_{i}^{1}}\mathrm{%
Tr}\left( \Phi _{i}\wedge F_{i}\right) +\frac{\theta }{g_{s}}\sum_{i}\int_{%
\mathbb{P}_{i}^{1}}\mathrm{Tr}\Phi _{i}\wedge \mathbf{\omega }+\frac{p_{i}}{%
2g_{s}}\int_{S_{i}^{2}}\mathrm{Tr}\Phi _{i}^{2}  \notag \\
&&-\frac{1}{g_{s}}\sum_{i\neq j}\int_{\mathbb{P}_{i}^{1}\cap \mathbb{P}%
_{j}^{1}}\mathrm{Tr}\left( \Phi _{ij}\wedge F_{ji}\right) +\sum_{i\neq j}%
\frac{p_{ij}}{2g_{s}}\int_{\mathbb{P}_{i}^{1}\cap \mathbb{P}_{j}^{1}}\mathrm{%
Tr}\left( \Phi _{ij}\Phi _{ji}\right) ,  \label{cher}
\end{eqnarray}%
where we have added the typical mass deformations $\frac{p_{i}}{2g_{s}}%
\mathrm{Tr}\Phi _{i}^{2}$ and by analogy $\frac{p_{ij}}{2g_{s}}\mathrm{Tr}%
\left( \Phi _{ij}\Phi _{ji}\right) $ with some $p_{ij}$ integers which a
priori should be related to the $p_{i}$ degrees of the line bundle.
Integrating out the scalar fields, one ends with%
\begin{equation}
\mathcal{S}_{2d}=-\sum_{i}\int_{\mathbb{P}_{i}^{1}}\frac{1}{2g_{i}^{2}}%
\mathrm{Tr}\left( F_{i}^{2}\right) -\sum_{i}\frac{\theta _{i}^{YM}}{g_{i}^{2}%
}\int_{\mathbb{P}_{i}^{1}}\mathrm{Tr}F_{i}-\sum_{i\neq j}\int_{\mathbb{P}%
_{i}^{1}\cap \mathbb{P}_{j}^{1}}\frac{1}{2G_{ij}^{2}}\mathrm{Tr}\left(
F_{ij}F_{ji}\right)  \label{sto}
\end{equation}%
where $g_{i}^{2}$ and $\theta _{i}^{YM}$ are as in (\ref{ts}) and where%
\begin{equation}
G_{ij}^{2}=\frac{p_{ij}g_{s}}{r}vol\left( \mathbb{P}_{i}^{1}\cap \mathbb{P}%
_{j}^{1}\right).
\end{equation}%
Note that for the base $\Delta _{n+1}$ realizing the elliptic curve in terms
of intersecting 2-spheres, the intersection $\mathbb{P}_{i}^{1}\cap \mathbb{P%
}_{j}^{1}$ is given by a finite and discrete set of points $P_{ij}$ of $%
\Delta _{n+1}$. These points have zero volumes $vol\left( \mathbb{P}%
_{i}^{1}\cap \mathbb{P}_{j}^{1}\right) $ and so
\begin{equation}
G_{ij}^{2}\simeq 0.
\end{equation}%
In the present case where $\Delta _{n+1}$ is taken as $\sum_{i}\mathbb{P}%
_{i}^{1}$, we have $\left( n+1\right) $ intersection points $P_{i,i+1}$. The
non zero intersection numbers is between neighboring spheres $\mathbb{P}%
_{i}^{1}$ and $\mathbb{P}_{i\pm 1}^{1}$,
\begin{equation}
\left( \left[ \mathbb{P}_{i}^{1}\right] \cap \left[ \mathbb{P}_{j}^{1}\right]
\right) =\delta _{j,i\pm 1}.
\end{equation}%
Implementing this specific data, the last term of eq(\ref{sto}) reduces then
to a sum of integrals over the following field densities
\begin{equation}
\frac{1}{2G_{i}^{2}}\mathrm{Tr}\left( \left\vert F_{i,i+1}\right\vert
^{2}\right) ,\qquad F_{n,n+1}=F_{n,0},
\end{equation}%
which diverge as long as $\left\vert F_{i,i+1}\right\vert ^{2}\neq 0$. This
property is not strange and was in fact expected. It has the behavior of a
Dirac function one generally use for implementing insertions. To exhibit
this feature, denote by $\mathbb{P}_{i}^{1}\cap \mathbb{P}_{i+1}^{1}=\left\{
P_{i}\right\} $, the points where the 2-spheres intersect. Then we have%
\begin{equation}
\int_{\mathbb{P}_{i}^{1}\cap \mathbb{P}_{i+1}^{1}}\frac{1}{2G_{i}^{2}}%
\mathrm{Tr}\left( \left\vert F_{i,i+1}\right\vert ^{2}\right) =\int_{\mathbb{%
P}_{i}^{1}}\frac{1}{2G_{i}^{2}}\delta \left( P-P_{i}\right) \mathrm{Tr}%
\left( \left\vert F_{i,i+1}\right\vert ^{2}\right)
\end{equation}%
where $\delta \left( P-P_{i}^{+}\right) $ is a Dirac delta function.
Combining the above results, one ends with the follwing field action of the
2d bosonic quiver gauge theory describing the brane configuration on the non
compact 4-cycle $\left[ \mathcal{D}_{4}\right] =\mathcal{O}(-p_{0},..,-p_{n})%
\rightarrow \Delta _{n+1}$ of the local CY$^{3}$,%
\begin{equation}
\mathcal{S}_{\Delta _{n}}=\sum_{i=0}^{n}\int_{\mathbb{P}_{i}^{1}}\left(
\frac{1}{2g_{i}^{2}}\mathrm{Tr}\left( F_{i}^{2}\right) +\frac{\theta
_{i}^{YM}}{g_{i}^{2}}\mathrm{Tr}F_{i}+\frac{1}{2G_{i}^{2}}\delta \left(
P-P_{i}\right) \mathrm{Tr}\left( \left\vert F_{i,i+1}\right\vert ^{2}\right)
\right).  \label{del}
\end{equation}%
In this relation, the coupling constants $g_{i}^{2}$ and $G_{i}^{2}$ are
expressed in terms of the string coupling $g_{s}$, the Kahler moduli of the
2-spheres of the base $\Delta _{n+1}$ and the degrees of the fiber $%
\mathcal{O}(-p_{0},..,-p_{n})$. The $F_{i}$'s are the $U\left( N_{i}\right) $ gauge
field strengths and
\begin{equation}
\delta \left( P-P_{i}\right) \times F_{i,i+1}
\end{equation}%
are insertion operators in bi-fundamental representations and are needed to
glue the spheres. The last term may be rewritten in different forms. For
example like $\sum_{i=0}^{n}\frac{1}{2G_{i}^{2}}\times $ $\mathrm{Tr}\left(
\left\vert F_{i,i+1}\left( P_{i}\right) \right\vert ^{2}\right) $ and it
depends on the $\mathbb{P}_{i}$'s.

\subsubsection{Path integral measure in 2d q-deformed quiver gauge theory}

Here we want to study the structure of the measure in the path integral
description of the partition function of the quiver gauge field action $%
\mathcal{S}_{2d}$ eq(\ref{cher},\ref{del}) which we rewrite as
\begin{eqnarray}
\mathcal{S}_{2d} &=&\sum_{i}\frac{p_{i}}{2g_{s}}\int_{\mathbb{P}_{i}^{1}}%
\mathrm{Tr}\Phi _{i}^{2}+\frac{1}{g_{s}}\sum_{i}\int_{\mathbb{P}_{i}^{1}}%
\mathrm{Tr}\left( \Phi _{i}\wedge F_{i}\right) +\sum_{i}\frac{\theta
_{i}^{YM}}{g_{s}}\int_{\mathbb{P}_{i}^{1}}\mathrm{Tr}\Phi _{i}\wedge \mathbf{%
\omega }  \notag \\
&&+\sum_{i}\frac{p_{ij}}{2g_{s}}\int_{\mathbb{P}_{i}^{1}\cap \mathbb{P}%
_{i+1}^{1}}\mathrm{Tr}\left( \Phi _{i,i+1}\Phi _{i+1,i}\right) -\frac{1}{%
g_{s}}\sum_{i}\int_{\mathbb{P}_{i}^{1}\cap \mathbb{P}_{i+1}^{1}}\mathrm{Tr}%
\left( \Phi _{i,i+1}\wedge F_{i+1,i}\right). \label{dal}
\end{eqnarray}%
We will give arguments indicating that bi-fundamentals contribute as well to
the deformation of path integral measure and in a very special manner. More
precisely, we give an evidence that adjoints and bi-fundamentals altogether
deform the measure by the quantity
\begin{equation}
\mathcal{J}_{\left( q_{0},..,q_{n}\right) }\left( \Phi \right)
=\dprod\limits_{i,j=0}^{n}\dprod\limits_{a_{i}<b_{i}=1}^{N_{i}}\left( \sqrt{%
\left[ \Phi _{a_{i}}-\Phi _{b_{i}}\right] _{q_{i}}\left[ \Phi _{a_{j}}-\Phi
_{b_{j}}\right] _{q_{j}}}\right) ^{-I_{ij}}.
\end{equation}%
In this relation, the $\Phi _{a_{i}}$'s are as in eq(\ref{be}) and where $%
I_{ij}$ is the intersection matrix of the 2-spheres of the $\Delta _{n+1}$
base. It is equal to minus the generalized Cartan matrix of affine $\hat{A}%
_{n}$.

To begin recall that the partition function $\mathcal{Z}_{YM}\left( \Sigma
_{g}\right) $ of topological 2d q-deformed $U\left( M\right) $ YM on a genus
$g$-Riemann surface $\Sigma _{g}$ is given by
\begin{equation}
\mathcal{Z}_{YM}\left( \Sigma _{g}\right) =\frac{1}{M!}\int^{\prime }\left(
\dprod\limits_{a=1}^{M}\left[ D\phi _{a}\right] \right) \left[ \Delta
_{H}\left( \phi _{a}\right) \right] ^{2-2g}\dprod%
\limits_{b=1}^{M}e^{g_{YM}^{-2}\int_{\Sigma _{g}}\left( \phi
_{b}F_{b}+\theta \phi _{b}+\frac{p}{2}\phi _{b}^{2}\right) }.  \label{vi}
\end{equation}%
where the $\phi _{a}$'s are the diagonal values of $U\left( M\right) $
unitary gauge symmetry. In this relation, $\Delta _{H}\left( \phi
_{a}\right) $ is given by,%
\begin{equation}
{\Large \Delta }_{H}\left( \phi \right) =\dprod\limits_{a<b=1}^{M}\left[
2\sin \left( \frac{\phi _{a}-\phi _{b}}{2}\right) \right] ,
\end{equation}
and is invariant under the periodic changes (\ref{pe}). It can take the
following form
\begin{equation}
\left[ {\Large \Delta }_{H}\left( \phi \right) \right] ^{2g-2}=\dprod%
\limits_{a<b=1}^{M}\left( \left[ x_{ab}\right] _{q}\right) ^{2-2g}
\end{equation}%
with%
\begin{equation}
x_{ab}=\frac{2i\left( \phi _{a}-\phi _{b}\right) }{g_{s}},\qquad \left[ x%
\right] _{q}=\left( q^{\frac{x}{2}}-q^{-\frac{x}{2}}\right) ,\qquad
q=e^{-g_{s}}.
\end{equation}%
Using\ this relation, we see that, on each 2-sphere $S_{i}^{2}$ of $\Delta
_{n+1}$, the correction to the path integral measure is
\begin{equation*}
\dprod\limits_{a_{i}<b_{i}=1}^{N_{i}}\left[ 2\sin \left( \frac{\phi
_{a_{i}}-\phi _{b_{i}}}{2}\right) \right] ^{2}
\end{equation*}%
which by setting $q_{i}=\exp \left( -g_{s}\frac{r_{i}}{r}\right) $ can be
put in the equivalent form
\begin{equation}
\mathcal{J}_{S_{i}^{2}}=\dprod\limits_{a_{i}<b_{i}=1}^{N_{i}}\left( \left[
\Phi _{a_{i}}-\Phi _{b_{i}}\right] _{q_{i}}\right) ^{2},\qquad i=0,...,n.
\end{equation}%
The power 2 in the right hand of above relation can be interpreted in terms
of the entries of the intersection matrix $I_{ii}=-K_{ii}$ of the \textit{%
i-th} 2-spheres of $\Delta _{n+1}$. This property is visible on eq(\ref{vi}%
) where the power $2-2g$ (Euler characteristics) is just the self-intersection
 of the Riemann surface $\Sigma _{g}$. As such, the above
relation can be put in the form
\begin{equation}
\mathcal{J}_{S_{i}^{2}}=\dprod\limits_{a_{i}<b_{i}=1}^{N_{i}}\left( \sqrt{%
\left[ \Phi _{a_{i}}-\Phi _{b_{i}}\right] _{q_{i}}\left[ \Phi _{a_{i}}-\Phi
_{b_{i}}\right] _{q_{i}}}\right) ^{K_{ii}},\,\quad K_{ii}=2,\,i=0,...,n.
\end{equation}%
This relation is very suggestive, it lets understand that this feature is a
special property of a more general situation where appears number 
intersection. More precisely, the structure of the deformation of the path
integral measure for local Calabi-Yau threefolds with some base B made of
2-cycles $\mathcal{C}_{i}$ with intersection matrix $I_{ij}=\left[ \mathcal{C}_{i}\right] .\left[
\mathcal{C}_{j}\right] $ should be as follows
\begin{equation}
\mathcal{J}_{B}=\dprod\limits_{i,j}\dprod\limits_{a_{i}<b_{i}=1}^{N_{i}}%
\left( \sqrt{\left[ \Phi _{a_{i}}-\Phi _{b_{i}}\right] _{q_{i}}\left[ \Phi
_{a_{j}}-\Phi _{b_{j}}\right] _{q_{j}}}\right) ^{-I_{ij}}.  \label{in}
\end{equation}%
In our concern, the 2d manifold is given by the base manifold $\Delta _{n+1}$
of the local Calabi-Yau threefold. In this case the intersection matrix of
the 2-cycles is given by  $I_{ij}=-K_{ij}$. So the partition function of the q-deformed
2d quiver gauge theory reads in general as%
\begin{equation}
\mathcal{Z}_{Quiver}=\int^{\prime }\left(
\dprod\limits_{i=0}^{n}\dprod\limits_{a_{i}=1}^{N_{i}}\frac{\left[ D\phi
_{a_{i}}\right] }{N_{i}!}\right) \left(
\dprod\limits_{i,j=0}^{n}\dprod\limits_{a_{i}<b_{i}=1}^{N_{i}}\left( \sqrt{%
\left[ \Phi _{a_{i}}-\Phi _{b_{i}}\right] _{q_{i}}\left[ \Phi _{a_{j}}-\Phi
_{b_{j}}\right] _{q_{j}}}\right) ^{K_{ij}}\right) e^{-S_{{\small Quiver}}}
\end{equation}%
where $S_{{\small Quiver}}$ is the action given by eq(\ref{del}-\ref{dal}).
Of course, here $\Delta _{n+1}$ is an elliptic curve and so one should have $%
\mathcal{J}_{B}=1$. This condition can be turned around and used rather as a
consistency condition to check the formula (\ref{in}). Indeed, for the case
of the 2-torus $T^{2}=S^{1}\times S^{1}$, we know that
\begin{equation}
\left[ T^{2}\right] .\left[ T^{2}\right] =0,
\end{equation}%
and so no q-deformation in agreement with eq(\ref{vi}). The same property is
valid for $\left[ \Delta _{n+1}\right] .$ But at this level, one may ask
what is then the link between the two realisations $T^{2}=S^{1}\times S^{1}$
and $\left[ \Delta _{n+1}\right] =\sum_{j=0}^{n}\left[ S_{j}^{2}\right] $.
The answer is that in the second case the role of the condition
\begin{equation}
\left( 2-2g\right) =0
\end{equation}%
that is obeyed by $S^{1}\times S^{1}$ is now played by the vanishing
property,%
\begin{equation}
\sum_{j=0}^{n}K_{ij}s_{j}=0
\end{equation}%
for affine Kac Moody algebras (with $s_{j}=1$ for affine $\widehat{A}_{n}$).
Let us check that $\mathcal{J}_{\Delta _{n+1}}$ eq(\ref{in}) is indeed equal
to unity. We will do it in two ways: \newline
First consider the simplest case given by the superconformal model with
gauge symmetry as in eq(\ref{at}) and specify the Kahler parameters at the
moduli space point where all the 2-spheres have the same area ($r_{i}=\frac{r%
}{n+1}$). In this case the quantity $\left[ \Phi _{a_{i}}-\Phi _{b_{i}}%
\right] _{q_{i}}$ is independent of the details of $\Delta _{n+1}$\ and so
the above formula reduces to,
\begin{equation}
\mathcal{J}_{\Delta _{n+1}}^{SCFT}\left( \Phi \right) =\left(
\dprod\limits_{a<b=1}^{M}\left[ \Phi _{a}-\Phi _{b}\right] _{q}\right)
^{\sum_{i=0}^{n}\sum_{j=0}^{n}K_{ij}}
\end{equation}%
which is equal unity ($\mathcal{J}\left( \Phi \right) =1$) due to the
relation $\sum_{j=0}^{n}K_{ij}=0$. In the general case where the gauge group
factors are arbitrary and for generic points in the moduli space, the
identity (\ref{in}) holds as well due to the same reason. For the
instructive case $n=2$ for instance, we have
\begin{eqnarray}
\mathcal{J}_{\Delta _{3}}^{QFT}\left( \Phi \right) &=&\left(
\dprod\limits_{a_{0}<b_{0}=1}^{N_{0}}\left[ \Phi _{a_{0}}-\Phi _{b_{0}}%
\right] _{q_{0}}\right) ^{2}\left( \dprod\limits_{a_{1}<b_{1}=1}^{N_{1}}%
\left[ \Phi _{a_{1}}-\Phi _{b_{1}}\right] _{q_{1}}\right) ^{2}\left(
\dprod\limits_{a_{2}<b_{2}=1}^{N_{2}}\left[ \Phi _{a_{2}}-\Phi _{b_{2}}%
\right] _{q_{2}}\right) ^{2}  \notag \\
&&\times \left( \dprod\limits_{a_{0}<b_{0}=1}^{N_{0}}\left[ \Phi
_{a_{0}}-\Phi _{b_{0}}\right] _{q_{0}}\dprod\limits_{a_{1}<b_{1}=1}^{N_{1}}%
\left[ \Phi _{a_{1}}-\Phi _{b_{1}}\right] _{q_{1}}\right) ^{-\frac{2}{2}}
\notag \\
&&\times \left( \dprod\limits_{a_{1}<b_{1}=1}^{N_{1}}\left[ \Phi
_{a_{1}}-\Phi _{b_{1}}\right] _{q_{1}}\dprod\limits_{a_{2}<b_{2}=1}^{N_{2}}%
\left[ \Phi _{a_{2}}-\Phi _{b_{2}}\right] _{q_{2}}\right) ^{-\frac{2}{2}}
\label{exa} \\
&&\times \left( \dprod\limits_{a_{2}<b_{2}=1}^{N_{2}}\left[ \Phi
_{a_{2}}-\Phi _{b_{2}}\right] _{q_{2}}\dprod\limits_{a_{0}<b_{0}=1}^{N_{0}}%
\left[ \Phi _{a_{0}}-\Phi _{b_{0}}\right] _{q_{0}}\right) ^{-\frac{2}{2}}.
\notag
\end{eqnarray}%
As we see, the diagonal terms of the first line of the right hand side of the
above relation are compensated by the off diagonal terms. Thus $\mathcal{J}%
_{\Delta _{3}}^{QFT}\left( \Phi \right) $ reduces exactly to unity and so
the 2d quiver gauge theory is not deformed. Nevertheless, one should keep in
mind that this would be a special property of a general result for 4d black
holes obtained from BPS D-branes in type IIA superstring  moving  on the follwing
general local Calabi-Yau threefolds
\begin{equation}
\mathcal{O}\left( \mathbf{m}\right) \oplus \mathcal{O}\left( \mathbf{-m-2}%
\right) \rightarrow \mathcal{B}_{k}.  \label{ba}
\end{equation}%
Here $\mathbf{m=}\left( m_{1},...,m_{k}\right) $ is an integer vector and $%
\mathcal{B}_{k}$ is a complex one dimension base consisting of the
intersection of $k$  2-spheres $S_{i}^{2}$ with some intersection matrix $%
I_{ij} $. Using Vinberg theorem \textrm{\cite{k,6,bs,mbs}}, the possible
matrices $I_{ij}$ may be classified basically into three categories. In the
language of Kac-Moody algebras, these correspond to: (\textbf{i}) Cartan
matrices of finite dimensional Lie algebras satisfying
\begin{equation}
\sum_{j}I_{ij}u_{j}>0
\end{equation}%
for some positive integer vector $\left( u_{j}\right) $. In this case the
resulting 2d quiver gauge theory is q-deformed. This theory has been also
studied in \cite{8}. (\textbf{ii}) Cartan matrices for affine Kac-Moody
algebras including simply laced ADE ones%
\begin{equation}
\sum_{j}I_{ij}u_{j}=0,
\end{equation}%
where now the $u_{j}^{\prime }s$\ are just the Dynkin weights. In this case,
the 2d quiver gauge theory is un-deformed due to the identity $%
\sum_{j}I_{ij}s_{j}=0$ where the $s_{j}^{\prime }$s are the Dynkin weights. (%
\textbf{iii}) Cartan matrices for indefinite Kac-Moody algebras where the
intersection matrix satisfies the condition
\begin{equation}
\sum_{j}I_{ij}u_{j}<0,
\end{equation}%
for some positive integer vector $\left( u_{j}\right) $. Here the 2d quiver
gauge theory is q-deformed.

\section{Conclusion}

\qquad In this paper, we have studied 4d black holes in type IIA superstring
theory on a particular class of local Calabi-Yau threefolds with compact
base made up of intersections of several 2-spheres. This study aims to test
OSV conjecture for the case of stacks of D-brane configurations on CY$^{3}$
cycles involving q-deformed 2d quiver gauge theories with gauge symmetry $G$
having more than one $U\left( N_{i}\right) $ gauge group factor. The class
of local threefolds we have considered with details is given by $%
X_{3}^{\left( \mathbf{m,-m-2,2}\right) }=$ $\mathcal{O}\left( \mathbf{m}%
\right) \oplus \mathcal{O}\left( \mathbf{-m-2}\right) \rightarrow \Delta
_{n+1}$ where $\mathbf{m}$ is a $\left( n+1\right) $ integer vector $\left(
m_{0},...,m_{n}\right) $. The $m_{i}$ components capture the non trivial
fibration of rank 2 line bundles of the local CY$^{3}$. They also define the
$U^{n+1}\left( 1\right) $ charges of the corresponding chiral superfields in
the supersymmetric gauged linear sigma model field realization. The compact
elliptic curve $\Delta _{n+1}$ is generally given by $\left( n+1\right) $
intersecting spheres according to affine Dynkin diagrams.

\qquad This study has been illustrated in the case of the local affine $\hat{%
A}_{n}$ model; but may a priori be extended to the other affine models
especially for DE simply laced series and beyond. Black holes in four
dimensions are realized by using D-brane configurations in type IIA
superstring compactified on $X_{3}^{\left( \mathbf{m,-m-2,2}\right) }$. The
topological twisted gauge theory on D4-brane wrapping 4-cycles in the local
CY$^{3}$s is shown to be reduced down to a 2d quiver gauge theory on the
base $\Delta _{n+1}$ and agrees with OSV conjecture. This agreement is
ensured by the results on $\mathcal{O}(2g+m-2)\oplus \mathcal{O}(-m)\rightarrow \Sigma _{g}$
obtained in \textrm{\cite{02,2}}. It is interestingly remarkable that
bi-fundamentals and adjoint scalars contribute to the deformed path integral
measure with opposite powers and compensate in the case of affine geometries
as shown in the example (\ref{exa}).

\qquad In developing this analysis, we have taken the opportunity to
complete partial results on the 2d $\mathcal{N}=2$ supersymmetric gauged
linear sigma model realization of the resolution of affine singularities and
the local Calabi-Yau threefold with non trivial fibrations. We have also
given comments on other black hole models testing OSV conjecture. They
concern the class of 4d black holes with D-branes wrapping cycles in local
threefolds with complex one dimension base manifolds $\mathcal{B}_{k}$ (\ref%
{ba}) classified by Vinberg theorem \cite{k}. The latter is known to
classify Kac-Moody algebras in three main sets: (i) ordinary finite
dimensional, (ii) affine Kac-Moody and (iii) indefinite set.

In the end, we would like to note that the computation given here can be
also done by using the topological vertex method. Aspects of this approach
have been discussed succinctly in present study. More details on this
powerful method as well as other features related to 2d quiver gauge
theories and topological string will be considered elsewhere.

\begin{acknowledgement}
:\qquad {\small This research work has been done in several steps, at
Rabat-Morocco, at ICTP- Italy, and at Faculdad de Ciencias, Universidad de
Zaragoza, Spain. AB and EHS would like to thank Departamento de Fisica
Teorica, Zaragoza for kind hospitality. They also thank Manuel Azorey, Luis.
J. Boya, Jose L. Cortes, Pablo Diaz, Sergio Montagnez and Antonio Segui for
fruitful discussions. AB is supported by Ministerio de Educacion y Cienca,
Spain, under grant FPA 2006-02315. LBD thanks "Le programme de la bourse
d'excellence, CNRST, Rabat". EHS would like to thank ICTP for hospitality
and Senior Associate Scheme for kind generosity. He thanks also N. Chair for
discussions. This research work is supported by the program Protars III
D12/25.}
\end{acknowledgement}

\end{document}